\documentclass[reprint,preprintnumbers,amsmath,amssymb,superscriptaddress,nofootinbib,preprintnumbers,floatfix,aps,prc]{revtex4-1}

\usepackage{graphicx}
\usepackage{dcolumn}

\usepackage{amsmath}
\usepackage{amssymb}
\usepackage{lineno}
\usepackage{isotope}
\usepackage{hyperref}
\usepackage{multirow}

\usepackage[position=bottom, caption=false]{subfig}

\newcommand{\ve}[1]{\ensuremath{\mathbf{#1}}}
\def\p{\mbox{\boldmath $p$}}
\def\q{\mbox{\boldmath $q$}}
\def\P{\mbox{{\bf p}$^\prime$}}

\hypersetup{
    colorlinks=true, 
    linkcolor=blue,  
   citecolor=red
}

\begin{document}

\title{Measurement of the \texorpdfstring{Ar$(e,e' p)$}{Ar(e,e' p)} and \texorpdfstring{Ti$(e,e' p)$}{Ti(e,e' p)} cross sections in Jefferson Lab Hall A}

\author{L.~Gu} \affiliation{Center for Neutrino Physics, Virginia Tech, Blacksburg, Virginia 24061, USA}
\author{D.~Abrams} \affiliation{Department of Physics, University of Virginia, Charlottesville, Virginia 22904, USA}
\author{A.~M.~Ankowski} \affiliation{SLAC National Accelerator Laboratory, Stanford University, Menlo Park, California 94025, USA}
\author{L. Jiang} \affiliation{Center for Neutrino Physics, Virginia Tech, Blacksburg, Virginia 24061, USA}
\author{B.~Aljawrneh} \affiliation{North Carolina Agricultural and Technical State University, Greensboro, North Carolina 27401, USA}
\author{S.~Alsalmi} \affiliation{King Saud University, Riyadh 11451, Kingdom of Saudi Arabia}
\author{J.~Bane} \affiliation{The University of Tennessee, Knoxville, Tennessee 37996, USA}
\author{A.~Batz} \affiliation{College of William and Mary, Williamsburg, Virginia 23187, USA}
\author{S.~Barcus} \affiliation{The College of William and Mary, Williamsburg, Virginia 23187, USA}
\author{M.~Barroso} \affiliation{Georgia Institute of Technology, Georgia 30332, USA}
\author{O.~Benhar} \affiliation{INFN and Dipartimento di Fisica, Sapienza Universit\`{a} di Roma, I-00185 Roma, Italy}
\author{V.~Bellini} \affiliation{INFN, Sezione di Catania, Catania, 95123, Italy}
\author{J.~Bericic} \affiliation{Thomas Jefferson National Accelerator Facility, Newport News, Virginia 23606, USA}
\author{D.~Biswas} \affiliation{Hampton University, Hampton, Virginia 23669, USA}
\author{A.~Camsonne} \affiliation{Thomas Jefferson National Accelerator Facility, Newport News, Virginia 23606, USA}
\author{J.~Castellanos} \affiliation{Florida International University, Miami, Florida 33181, USA}
\author{J.-P.~Chen} \affiliation{Thomas Jefferson National Accelerator Facility, Newport News, Virginia 23606, USA}
\author{M.~E.~Christy} \affiliation{Hampton University, Hampton, Virginia 23669, USA}
\author{K.~Craycraft} \affiliation{The University of Tennessee, Knoxville, Tennessee 37996, USA}
\author{R.~Cruz-Torres} \affiliation{Massachusetts Institute of Technology, Cambridge, Massachusetts 02139, USA}
\author{H.~Dai} \affiliation{Center for Neutrino Physics, Virginia Tech, Blacksburg, Virginia 24061, USA}
\author{D.~Day} \affiliation{Department of Physics, University of Virginia, Charlottesville, Virginia 22904, USA}
\author{S.-C.~Dusa} \affiliation{Thomas Jefferson National Accelerator Facility, Newport News, Virginia 23606, USA}
\author{E.~Fuchey} \affiliation{University of Connecticut, Storrs, Connecticut 06269, USA}
\author{T.~Gautam} \affiliation{Hampton University, Hampton, Virginia 23669, USA}
\author{C.~Giusti} \affiliation{Dipartimento di Fisica, Universit\`{a} degli Studi di Pavia and INFN, Sezione di Pavia,  I-27100 Pavia, Italy}
\author{J.~Gomez} \affiliation{Thomas Jefferson National Accelerator Facility, Newport News, Virginia 23606, USA}
\author{C.~Gu} \affiliation{Duke University, Durham, North Carolina 27708, USA}
\author{T.~Hague} \affiliation{Kent State University, Kent, Ohio 44242, USA}
\author{J.-O.~Hansen} \affiliation{Thomas Jefferson National Accelerator Facility, Newport News, Virginia 23606, USA}
\author{F.~Hauenstein} \affiliation{Old Dominion University, Norfolk, Virginia 23529, USA}
\author{D.~W.~Higinbotham} \affiliation{Thomas Jefferson National Accelerator Facility, Newport News, Virginia 23606, USA}
\author{C.~Hyde} \affiliation{Old Dominion University, Norfolk, Virginia 23529, USA}
\author{C.~Keppel} \affiliation{Thomas Jefferson National Accelerator Facility, Newport News, Virginia 23606, USA}
\author{S.~Li} \affiliation{University of New Hampshire, Durham, New Hampshire 03824, USA}
\author{R.~Lindgren} \affiliation{Department of Physics, University of Virginia, Charlottesville, Virginia 22904, USA}
\author{H.~Liu} \affiliation{Columbia University, New York, New York 10027, USA}
\author{C.~Mariani}\email{mariani@vt.edu} \affiliation{Center for Neutrino Physics, Virginia Tech, Blacksburg, Virginia 24061, USA}
\author{R.~E.~McClellan} \affiliation{Thomas Jefferson National Accelerator Facility, Newport News, Virginia 23606, USA}
\author{D.~Meekins} \affiliation{Thomas Jefferson National Accelerator Facility, Newport News, Virginia 23606, USA}
\author{R.~Michaels} \affiliation{Thomas Jefferson National Accelerator Facility, Newport News, Virginia 23606, USA}
\author{M.~Mihovilovic} \affiliation{Jozef Stefan Institute, Ljubljana 1000, Slovenia}
\author{M.~Murphy} \affiliation{Center for Neutrino Physics, Virginia Tech, Blacksburg, Virginia 24061, USA}
\author{D.~Nguyen} \affiliation{Department of Physics, University of Virginia, Charlottesville, Virginia 22904, USA}
\author{M.~Nycz} \affiliation{Kent State University, Kent, Ohio 44242, USA}
\author{L.~Ou} \affiliation{Massachusetts Institute of Technology, Cambridge, Massachusetts 02139, USA}
\author{B.~Pandey} \affiliation{Hampton University, Hampton, Virginia 23669, USA}
\author{V.~Pandey} \altaffiliation{Present Address: Department of Physics, University of Florida, Gainesville, FL 32611, USA}
\affiliation{Center for Neutrino Physics, Virginia Tech, Blacksburg, Virginia 24061, USA}
\author{K.~Park} \affiliation{Thomas Jefferson National Accelerator Facility, Newport News, Virginia 23606, USA}
\author{G.~Perera} \affiliation{Department of Physics, University of Virginia, Charlottesville, Virginia 22904, USA}
\author{A.~J.~R.~Puckett} \affiliation{University of Connecticut, Storrs, Connecticut 06269, USA}
\author{S.~N.~Santiesteban} \affiliation{University of New Hampshire, Durham, New Hampshire 03824, USA}
\author{S.~\v{S}irca} \affiliation{University of Ljubljana, Ljubljana 1000, Slovenia} \affiliation{Jozef Stefan Institute, Ljubljana 1000, Slovenia}
\author{T.~Su} \affiliation{Kent State University, Kent, Ohio 44242, USA}
\author{L.~Tang} \affiliation{Hampton University, Hampton, Virginia 23669, USA}
\author{Y.~Tian} \affiliation{Shandong University, Shandong, 250000, China}
\author{N.~Ton} \affiliation{Department of Physics, University of Virginia, Charlottesville, Virginia 22904, USA}
\author{B.~Wojtsekhowski} \affiliation{Thomas Jefferson National Accelerator Facility, Newport News, Virginia 23606, USA}
\author{S.~Wood} \affiliation{Thomas Jefferson National Accelerator Facility, Newport News, Virginia 23606, USA}
\author{Z.~Ye} \affiliation{Physics Division, Argonne National Laboratory, Argonne, Illinois 60439, USA}
\author{J.~Zhang} \affiliation{Department of Physics, University of Virginia, Charlottesville, Virginia 22904, USA}

\collaboration{The Jefferson Lab Hall A Collaboration}

\begin{abstract}

The E12-14-012 experiment, performed in Jefferson Lab Hall A, has collected exclusive electron-scattering data $(e, e'p)$ in parallel kinematics using natural argon and natural titanium targets. Here, we report the first results of the analysis of the data set corresponding to  beam energy 2,222~GeV,  electron scattering angle $21.5$ deg, and proton emission angle $-50$ deg. The differential cross sections, measured with $\approx$4\% uncertainty,  have been studied as a function of missing energy and missing momentum, and compared to the results of Monte Carlo simulations, obtained from a model based on the Distorted Wave Impulse Approximation.

\end{abstract}

\preprint{JLAB-PHY-21-3197}
\preprint{SLAC-PUB-17571}
%
%
\maketitle
\section{Introduction}
%
%
Jefferson Lab experiment E12-14-012 was primarily aimed at obtaining the proton spectral function (SF) of the nucleus \isotope[40]{Ar} from a measurement of the cross section of the $(e,e^\prime p)$ reaction
\begin{align}
\label{eep:reaction}
e + A \to e^\prime + p + (A-1)^*,
\end{align}
in which the scattered electron and the knocked out proton are detected in coincidence. Here $A$ denotes the target nucleus in its ground state, while the recoiling $(A-1)$-nucleon system can be either in the ground state or in any excited state.

Nucleon knockout processes have long been recognized as being ideally suited to study the momentum and removal energy distribution of protons bound in atomic nuclei~\cite{Benhar_NPN}. Compared to the pioneering studies carried out using proton beams, see, e.g., Ref.~\cite{p2p}, $(e,e' p)$ experiments have clear advantages, because they are largely unaffected by strong initial and final state interactions (FSI) between the beam particle and the target, and give access to the properties of deeply bound protons in medium-mass and heavy nuclei~\cite{Jacob:1962}.

Under the basic assumption that the scattering process involves individual nucleons, and neglecting FSI between the outgoing proton and the spectator nucleons, the momentum and removal energy of the knocked out particle, ${\bf p}$ and $E$, can be reconstructed from measured kinematical variables, and the cross section of the process is written in simple factorized form in terms of the spectral function of the target nucleus, $P({\bf p},E)$, trivially related to the nucleon Green's function, $G({\bf p},E)$, through
\begin{align}
\label{specfact}
P({\bf p},E) =  \frac{1}{\pi}\  {\rm Im} \ G({\bf p},E).
\end{align}
As a consequence, the spectral function---yielding the probability to remove a proton with momentum ${\bf p}$ from the
target nucleus leaving the residual system with excitation energy $E-E_{\rm thr}$, with $E_{\rm thr}$ being the proton emission threshold---can be readily obtained from the data.

Significant corrections to the somewhat oversimplified scheme outlined above\textemdash referred to as Plane Wave Impulse Approximation, or PWIA\textemdash arise from the occurrence of FSI. The large body of work devoted to the analysis of $(e,e^\prime p)$ data has provided convincing evidence that the effects of FSI can be accurately included by replacing the plane wave describing the motion of the outgoing proton with a distorted wave, eigenfunction of a phenomenological optical potential accounting for its interactions with the mean field of the residual nucleus. In general, the $(e,e^\prime p)$ cross section computed within this approach, known as Distorted Wave Impulse Approximation, or DWIA, involves the off-diagonal spectral function, and cannot be written in factorized form~\cite{Gross}.  However, an approximate procedure restoring factorization, referred to as factorized DWIA, has been shown to yield accurate results in the case of parallel kinematics, in which the momentum of the outgoing proton and the momentum transfer are parallel~\cite{Bof79}. In this kinematical setup, the spectral function can still be reliably obtained from $(e,e^\prime p)$ data after removing the effects of FSI.

Additional corrections to the PWIA arise from the distortion of the electron wave functions resulting from interactions with the Coulomb field of the nucleus. However, it has been shown that, for nuclei as heavy as \isotope[40][]{Ca},  this effect can be accurately taken into account using an effective momentum transfer~\cite{giu88}.

Systematic measurements of $(e,e^\prime p)$ cross sections in the kinematical regime in which the recoiling nucleus is left in a bound state, performed at Saclay~\cite{fru84} and NIKHEF-K~\cite{NIKHEF}, have allowed the determination of  the spectral functions of a broad set of nuclei. These studies have provided a wealth of information on the energies and momentum distributions of shell-model states belonging to the Fermi sea of the target nuclei, showing at the same time the limitations of the mean-field description and the importance of correlation effects~\cite{Benhar_NPN}.

Besides being a  fundamental quantity of nuclear many-body theory, containing important dynamical information, the spectral function is a powerful tool,  allowing to obtain the cross sections of a variety of nuclear scattering processes in the kinematical regime in which the beam particles primarily interact with individual nucleons, and FSI can be treated as corrections. Applications to inclusive electron-nucleus scattering have offered vast evidence that the formalism based on spectral functions provides a comprehensive and consistent framework for the calculation of nuclear cross sections in a broad kinematical region, extending from quasielastic (QE) scattering to resonance production and deep-inelastic scattering~\cite{BP,BFFS,ABS}.

Over the past several years, a great deal of work has been devoted to applying the spectral function formalism to the study of neutrino-nucleus interactions, whose quantitative understanding is needed for the interpretation of accelerator-based searches of neutrino oscillations, see, e.g., Refs.~\cite{PhysRep,Ankowski:2016jdd}.
In this context, it should be noted that the capability to describe a variety of reaction channels within a unified approach is a critical requirement, because the energy of the beam particles is distributed according to a broad flux, typically ranging from a few hundreds of MeV to a few GeV. Moreover, the knowledge of the spectral function greatly improves the accuracy of reconstruction of the neutrino energy, a key quantity in the oscillation analysis~\cite{BM,Jen:2014aja}.

Realistic models of the nuclear spectral functions have been obtained from the approach based on the local density approximation, or LDA, in which the information on the shell-model structure extracted from $(e,e^\prime p)$ data is combined to the results of accurate calculations of uniform nuclear matter at various densities~\cite{BFFS}.
The existing calculations of neutrino-nucleus cross sections employing LDA spectral functions~\cite{ABS,BM,PRD,BM2,Ankowski:2010yh,coletti,Ankowski:2011ei,Ankowski:2012ei,Ankowski:2013gha,Ankowski:2015ega, Ankowski:2015lma,Vagnoni:2017hll,Ankowski:2017yvm}, however, are limited to the isospin-symmetric $p$-shell targets \isotope[16][]{O} and \isotope[12][]{C}.
Therefore, the results of these studies are applicable to experiments using water-\v{C}erenkov detectors, e.g. Super-Kamiokande~\cite{Super-K:detector}, and mineral oil detectors, e.g. MiniBooNE~\cite{MiniBooNE:detector}.

The analysis of the data collected by the ongoing and future experiments using liquid-argon time-projection chambers, notably the Fermilab Short-Baseline Neutrino program (SBN)~\cite{SBNProposal:2015} and the Deep Underground Neutrino Experiment  (DUNE)~\cite{Abi:2020wmh}, will require the extension of this approach to the case of a heavier target with large neutron excess. Moreover, in DUNE the proton and neutron spectral functions will both be needed, to extract the Dirac phase $\delta_{CP}$ from a comparison of neutrino and antineutrino oscillations, and achieve an accurate description of pion production on protons and neutrons.

In the absence of direct measurements, information on the neutron momentum and removal energy distribution in $\isotope[40][18]{Ar}$ can be inferred from $\isotope[][]{\rm Ti}(e,e^\prime p)$ data, exploiting the correspondence between the proton spectrum of titanium, having charge $Z=22$, and the neutron spectrum of argon, having $A-Z = 22$.  The viability of this procedure is supported by the results of Ref.~\cite{BRS}, whose authors have performed a calculation of the inclusive $\isotope[40]{Ar}(e,e^\prime)$ and $\isotope[48]{Ti}(e,e^\prime)$ cross sections within the framework of the self-consistent Green's function approach. The aim of Jlab experiment E12-14-012, is the determination of the proton spectral functions of argon and titanium from the corresponding $(e,e^\prime p)$ cross sections. 

In this article, we present the first results of our analysis. In Sec.~\ref{sec:ExperimentalSetup} we discuss the kinematic setup, the detectors and their resolutions, and our definitions of signal and backgrounds. In Sec.~\ref{sec:data_analysis} we introduce the missing energy and the missing momentum, which are the fundamental variables of our analysis, and discuss the main elements of the Monte Carlo (MC) simulations employed for event simulation. Sec.~\ref{sec:Uncertainties} is devoted to the uncertainties associated with our analysis, while in Sec.~\ref{sec:comparison} the 
measured missing energy and missing momentum distributions are compared with the MC predictions. Finally, in Sec.~\ref{sec:Summary} we summarize our work and draw the conclusions.

\begin{table*}[tb]
\caption{\label{tab:kinematic}Kinematics settings used to collect the data analyzed here.}
\begin{ruledtabular}
\begin{tabular}{ c c c c c c c c c c c }
    & $E_{e}^\prime$ & $\theta_e$  & $Q^2$ & $|\P|$ & $T_{p^\prime}$ & $\theta_{p^\prime}$ & $|{\bf q}|$ & $p_m$ & $E_m$ \\
    & (GeV) & (deg) & (GeV$^2/c^2$) & (MeV/$c$) & (MeV) & (deg) &(MeV/$c$) & (MeV/$c$) & (MeV)\\
    \colrule
    {Ar}&  1.777  & 21.5 & 0.549 & 915 & 372  &  $-50.0$  & 865  &   50    & 73  \\
    {Ti}&  1.799  & 21.5 & 0.556 & 915 & 372  &  $-50.0$  & 857  &   58    & 51   \\
\end{tabular}
\end{ruledtabular}
\end{table*}

\section{Experimental Setup}\label{sec:ExperimentalSetup}
The experiment E12-14-012 was performed at Jefferson lab in Spring 2017. Inclusive $(e,e^\prime)$ and exclusive $(e,e^\prime p)$ electron scattering data were collected on targets of natural argon and natural titanium, as well as on calibration and background targets of carbon and aluminum. The average neutron numbers calculated according to the natural abundances of isotopes are 21.98 for
argon and 25.92 for titanium~\cite{Murphy:2019wed}. Therefore, from now on we will refer to the targets considered here as \isotope[40][]{Ar} and \isotope[48][]{Ti}, for brevity.

The E12-14-012 experiment used an electron beam of energy 2.222~GeV provided by the Continuous Electron Beam Accelerator Facility (CEBAF) at Jefferson Lab. The average beam current was approximately 15~$\mu$A for the \isotope[40][]{Ar} target and 20~$\mu$A for the \isotope[48][]{Ti} target. The scattered electrons were momentum analyzed and detected in the left high-resolution spectrometer (HRS) in Hall A and the coincident protons were similarly analyzed in the right HRS. The spectrometers are equipped with two vertical drift chambers (VDCs) providing tracking information~\cite{Fissum:2001st}, two scintillator planes for timing measurements and triggering, double-layered lead-glass calorimeter, a gas \v{C}erenkov counter used for particle identification~\cite{Alcorn:2004sb}, pre-shower and shower detectors (proton arm only)~\cite{Alcorn:2004sb} and pion rejectors (electron arm only)~\cite{Alcorn:2004sb}. The HRSs were positioned with the electron arm at central scattering angle $\theta_e=21.5$ deg and the proton arm at an angle $\theta_{p^\prime}=-50$ deg. The beam current and position, the latter being critical for the electron-vertex reconstruction and momentum calculation, were monitored by resonant radio-frequency cavities (beam current monitors, or BCMs~\cite{Alcorn:2004sb}) and cavities with four antennae (beam position monitors, or BPMs~\cite{Alcorn:2004sb}), respectively. The beam size was monitored using harp scanners, which consists of a thin wire which moves  through the beam. We used a raster of $2\times2$ mm$^2$ area to spread the beam and avoid overheating the target.

The experiment employed also an aluminum target and a set of carbon targets, used to evaluate backgrounds and monitor the spectrometers optics. The aluminum target was made of two identical foils of the Al-7075 alloy with a thickness of $0.889\pm0.002$~g/cm$^2$. One of the aluminum foils was positioned to match the entrance and the other to match the exit windows of the argon gas target cell. The two thick foils were separated by a distance of 25~cm, corresponding to the length of the argon gas cell and the Al foil's thickness.

The analysis presented here uses data collected with the settings given in Table~\ref{tab:kinematic}. All of our data were taken in parallel kinematics, in which the momentum transfer, $\q$, and  the momentum of the outgoing proton, $\P$, are parallel. The only difference of data collection setting for \isotope[40][]{Ar} and \isotope[48][]{Ti} is the scattered electron energy. %

\par The VDCs' tracking information was used to determine the momentum and to reconstruct the direction (in-plane and out-of-plane angles) of the scattered electron and proton, and to reconstruct the interaction vertex at the target. We used both the electron and proton arm information separately  to reconstruct the interaction vertex and found them in very good agreement. The transformation between focal plane and target quantities was computed using an optical matrix, the accuracy of which was verified using the carbon multi-foil target data and sieve measurements as described in previous papers~\cite{Dai:2018gch,Dai:2018xhi,Murphy:2019wed}. Possible variations of the optics and magnetic field in both HRSs are included in the analysis as systematic uncertainties related to the optics.

Several different components were used to build the triggers: the scintillator planes on both the electron and proton spectrometers, along with  signals from the gas \v{C}erenkov (GC) detector, the pion rejector (PR), the pre-shower and the shower detector (PS). Table~\ref{trigger_table} lists the trigger configurations, including details on how the signals from the various detector components are combined to form a trigger.

\begin{table}[htb!]
\caption{\label{trigger_table}Trigger lists detailing how the signals from different detector components are combined. LEFT and RIGHT identify the electron and proton arm, respectively.}
\begin{ruledtabular}
\begin{tabular}{ @{\hspace{3em}}c @{\qquad} l  @{\hspace{3em}}}
 T1  & $(S_0\&\&S_2)$ {\sc and} (GC$||$PR) [LEFT]	\\
       &  {\sc and} (S0\&\&S2) [RIGHT] 			\\
 T2  & $(S_0||S_2)$ {\sc and} (GC$||$PR) [LEFT]	\\
     &  {\sc and} $(S_0||S_2)$ {\sc and} {\sc not}(PS) [RIGHT]	\\
 T3  & $(S_0\&\&S_2)$ {\sc and} (GC$||$PR) [LEFT] 	\\
 T4  & $(S_0\&\&S_2)$ [RIGHT] 				\\
 T5  & $(S_0||S_2)$ {\sc and} (GC$||$PR) [LEFT] 	\\
 T6  & $(S_0||S_2)$ {\sc and} {\sc not}(PS) [RIGHT] 		\\
\end{tabular}
\end{ruledtabular}
\end{table}

The triggers used for identifying electron and proton coincidence events were T1 and T2, where T2 was used to provide a data sample to calculate the overall T1 trigger efficiency and we were able to compute the efficiency of T1 using also the product of T3 and T4 efficiencies. If the proton and electron observations from the same event were perfectly paired, these values would be the same as T1 trigger efficiency.

Electrons and protons were selected in their corresponding HRS requiring only one reconstructed good track. For the electron we required also an energy deposit of at least 30\% in the lead calorimeter (${E_\text{cal}}/{p} > 0.3$) and a signal in the \v{C}erekov detector of more than 400~analog-digital-converter (ADC) counts.  Furthermore, the tracks were required to be within $\pm$3~mrad of the in-plane angle and $\pm$6~mrad of the out-of-plane angle with respect to the center ray of the spectrometer and have a $dp/p$ of $\pm$0.06. Those latter conditions focused on removing events coming from the acceptance edges of the spectrometers. We used a cut on $\beta$ for the proton arm between 0.6 and 0.8 to further isolate protons. We only included in our analysis events  in which both the electron and the proton were recorded in a T1 trigger timing window and for which the difference in the start time of the individual triggers was of just few ns (time coincidence cut).  For the argon target we also required that the events originated within the central $\pm$10~cm of the target cell to exclude contamination from the target entry and exit windows. By measuring events from the thick Al foils, positioned at the same entry and exit window of the target, we determined that the target cell contributions to the measured cross section was negligible ($<$0.1\%). The same gas cell was used in another set of experiments and the contribution from an empty gas cell was measured and confirmed a very low contamination of events coming from the Al windows~\cite{Santiesteban:2018qwi}.
The spectrometer optics were calibrated using sieve slit measurements and their positions and angles were surveyed before and after moving the spectrometers for each kinematic settings. The survey precision was 0.01~rad and 0.01~mm respectively for the angle and positions of the spectrometers.
\par The efficiencies of the elements in the detector stack were studied by comparing rates in various combinations of secondary triggers as in Ref.~\cite{Dai:2018gch,Dai:2018xhi,Murphy:2019wed}. Table~\ref{tab:eff} summarizes the efficiency for the trigger, acceptances and kinematical cuts.
The live-time of the electronics was computed using the rates from scalers, which were independent of triggered events. The acceptance cuts efficiencies were computed using the MC simulation~\cite{Arrington:1999}.
The efficiency calculations that are based on MC were evaluated multiple times using slightly different SF models in the MC. The effect of theory models was found to be negligible. Our MC model contains nuclear transparency correction~\cite{Frankfurt:2001,Arrington:1999}, but does not account for all FSI effects. We have studied the role of FSI by looking at kinematical distributions for various MC samples obtained using different ranges of the missing momentum $p_m$, defined as in Eq.~\eqref{pmiss}, from lower to higher. We found that the electron arm $dp/p$ distributions showed slight variations. We then decided not to use the electron arm $dp/p$ as a kinematical cut in our analysis.
The trigger efficiencies were computed using the other available trigger as described above. The time coincidence cut efficiency was evaluated selecting a sample of more pure signal events (using a tighter $\beta$ cut) and looking at the ratio of events with and without the time coincidence cuts.
The overall efficiency (between 39.6\% and 48.9\% across all kinematic regions for the \isotope[40][]{Ar} target, and between 46.8\% and 48.1\% for the \isotope[48][]{Ti} target) includes cuts on the coincidence triggers, calorimeters, both the lead and the \v{C}erenkov counter, track reconstruction efficiency, live-time, tracking and $\beta$ cut.
\begin{table}
\caption{\label{tab:eff} Summary of the efficiency analysis for the argon and titanium targets.}
\begin{ruledtabular}
\begin{tabular}{@{}l c c c c@{}}
								        &		& Ar target         	    & Ti target \\
\hline				
\phantom{1. }a.~Live time 					& 		& 98.0\%     		    &    98.9\% \\
\phantom{1. }b.~Tracking 					&		& 98.3\%    		    &    98.3\% \\
\phantom{1. }c.~Trigger 					&		& 92.3\%   		    &    96.9\% \\
\phantom{1. }d.~\v{C}erenkov cut 			&		& 99.9\% 			    &    96.6\% \\
\phantom{1. }e.~Calorimeter cut 			&		& 97.8\% 			    &    98.1\% \\
\phantom{1. }f.~$\beta$ cut			    	&		& 95.6\% 			    &    95.3\% \\
\phantom{1. }g.~Coincidence time cut		&     		& 54.8\% 			    &    55.5\% \\
\end{tabular}
\end{ruledtabular}
\end{table}
\section{Data Analysis}\label{sec:data_analysis}
\subsection{The \texorpdfstring{$(e,e^\prime p)$}{(e,e^\prime p)} cross section}
In electron-nucleus scattering an incident electron, with energy $E_e$, is scattered from a nucleus of mass $M_A$ at rest. Electron scattering is generally described in the one-photon exchange approximation, according to which the incident electron exchanges a space-like photon, of energy $\omega$ and momentum $\q$, with the target nucleus.

In $(e,e^\prime p)$ experiments the scattered electron and a proton are detected in coincidence in the final state, and their momentum and energy are completely determined. If, in addition,  the kinematics is chosen such that  the residual nucleus is left in a specific bound state, the reaction is said to be exclusive.

In the following, $\P$,  $T_{p^\prime}$, and $M$ will denote the momentum, kinetic energy, and mass of the outgoing proton, while the corresponding quantities associated with the recoiling residual nucleus will be denoted ${\p}_R$, $T_R$, and $M_R$. The missing momentum and missing energy are obtained from the measured kinematical quantities using the definitions
%
\begin{equation}
\label{pmiss}
	\p_m = \q-\P= \p_R,
\end{equation}
and
\begin{equation}
\label{Emiss}
	E_m=\omega-T_{p^\prime}-T_R.
\end{equation}
Exploiting energy conservation, implying
\begin{equation}
\label{en:cons}
\omega + M_A = M + T_{p^\prime} + M_R + T_R,
\end{equation}
and writing the mass of the residual nucleus in the form
\begin{align}
M_R = M_A - M + E_{\rm thr} + E_x = M_{A-1} + E_x,
\end{align}
where $E_{\rm thr}$ and $M_{A-1}$ denote the proton emission threshold and the mass of $(A-1)$-nucleon system
in its ground state, respectively,  Eq.~\eqref{Emiss} can be rewritten
\begin{align}
E_m = E_{\rm thr} + E_x.
\end{align}

The usual description of the exclusive $(e,e^\prime p)$ reaction in the QE region assumes the direct knockout mechanism, which naturally emerges within the impulse approximation (IA). According to this picture, the electromagnetic probe interacts through a one-body current with the quasi-free knocked out proton, while all other nucleons in the target act as spectators. In addition, if FSI between the outgoing nucleon and the spectators is negligible, PWIA can be applied, and the $(e,e^\prime p)$ cross section reduces to the factorized form
\begin{equation}
	\frac{d^6 \sigma }{ d\omega d\Omega_{ e^\prime} d T_{p^\prime} d\Omega_{p^\prime}} = K \sigma_{ep} P(-\p_m,E_m),
\label{eq:PWIA_xs}
\end{equation}
where $K = |\P| E_{p^\prime}$, with $E_{p^\prime} = \sqrt{\P^2 + M^2}$. Here, $\sigma_{ep}$ is the differential cross section describing electron scattering off a bound moving proton, stripped of the flux factor and the energy conserving delta-function~\cite{deforest:PLB,deforest}, while $P(-\p_m,E_m)$ is the proton spectral function of Eq.~\eqref{specfact}. Note that Eqs.~\eqref{pmiss} and \eqref{Emiss} imply that the arguments of the spectral function can be identified with the initial momentum and the removal energy of the struck nucleon, respectively. Therefore, Eq.~\eqref{eq:PWIA_xs} shows that within PWIA the nuclear spectral function, describing the proton momentum and energy distribution of the target nucleus,
can be readily extracted from the measured $(e,e^\prime p)$ cross section.

When FSI are taken into account, and the outgoing proton is described by a distorted wave function as prescribed by DWIA, the initial momentum of the struck nucleon is not trivially related to the measured missing momentum, and the cross section can no longer be written as in Eq.~\eqref{eq:PWIA_xs}. However, the occurrence of $y$-scaling in inclusive electron-nucleus
scattering~\cite{yscaling1,yscaling2}\textemdash whose observation in the analysis of the Ar$(e,e^\prime)$ and Ti$(e,e^\prime)$ data is discussed in Refs.~\cite{Dai:2018gch,Dai:2018xhi}\textemdash indicates that the formalism based on factorization is still largely applicable in the presence of FSI.

In principle, within the approach of Refs.~\cite{bof93,bof96,bof82}, the bound and scattering states are both derived from an energy dependent non-Hermitian optical-model Hamiltonian. While being fully consistent, however, this treatment involves severe difficulties. In practice,  the bound-state proton wave functions are generally obtained from phenomenological approaches---although a~few studies based on realistic microscopic models of the nuclear Hamiltonian have been carried out for light and medium-heavy nuclei~\cite{radici, bisconti}---while the scattering states are eigenfunctions of  phenomenological optical potentials, the parameters of which are determined through a fit to elastic proton-nucleus scattering data.

The PWIA description provides a clear understanding of the mechanism driving the  $(e,e^\prime p)$ reaction, and the ensuing factorized expression of the coincidence cross section, Eq.~(\ref{eq:PWIA_xs}),  is essential to obtain from the data an intrinsic property of the target, such as the spectral function, independent of kinematics. As pointed out above, however, the occurrence of FSI leads to a violation of factorization,  and makes the extraction of the spectral function from the measured cross section more complicated~\cite{bof96,bern82}. Additional factorization-breaking corrections arise from the distortion of the electron wave functions, resulting from interactions with the Coulomb field of the target~\cite{udi93,giu87,giu88}.

The general conditions to recover a factorized expression of the cross section are discussed in Refs.~\cite{bof93,bof96,Bof79,bof80,vig04}. If these requirements are fulfilled, the DWIA cross section can be written in terms of a distorted spectral function according to
\begin{equation}
\frac{d^6 \sigma }{ d\omega d\Omega_{ e^\prime} d T_{p^\prime} d\Omega_{p^\prime}} = K \sigma_{ep} P^D(\P,-\p_m,E_m).
\label{eq:dsf}
\end{equation}
Note, however that, unlike the spectral function appearing in Eq.~\eqref{eq:PWIA_xs}, the distorted spectral function is {\em not} an intrinsic property of the target, because it depends explicitly on the momentum of the outgoing nucleon, which in turn depends on the momentum transfer. The most prominent effects of the inclusion of FSI within the framework of DWIA are a shift and a suppression of the missing momentum distributions, produced by the real and imaginary part of the optical potential, respectively.


\subsection{Data analysis details}\label{subsec:Analysis}

The measured cross sections are usually analyzed in terms of missing-energy and missing-momentum distributions. For a value of $E_m$ corresponding to a peak in the experimental missing-energy distribution, the data are usually presented in terms of the reduced cross section as a function of $p_m=|\p_m|$. The reduced cross section, obtained from the measured cross section dividing out the kinematic factor $K$ and the electron-proton cross section $\sigma_{ep}$ can be identified with the spectral function in PWIA and with the distorted spectral function in the factorized DWIA of Eq.~(\ref{eq:dsf}). The off-shell extrapolation of de Forest~\cite{, deforest:PLB,deforest} is generally used to describe the bound nucleon cross section.

The experimental reduced cross sections can be compared with the corresponding reduced cross section calculated using different theoretical models.
The comparison of the results obtained from the un-factorized and factorized approaches allows one to make an estimate of the accuracy of the factorization scheme, as well as the sensitivity to the different factorization-breaking contributions.

The six-fold differential cross section as a function of $p_{m}$ and $E_{m}$ was extracted from the data using the $(e,e^\prime p)$ event yield $Y$ for each $p_{m}$ and $E_{m}$ bin
\begin{equation}
\frac{d^6 \sigma }{ d\omega d\Omega_{ e^\prime} d T_{p^\prime} d\Omega_{p^\prime}}  = \frac{{Y}(p_{m},E_{m})}{B \times lt \times \rho \times BH \times V_B \times C_\text{rad}} ,
\label{eq:yield}
\end{equation}
where $B$ is the total accumulated beam charge, $lt$ is the live-time of the detector (fraction of time that the detector was able to collect and write data to disk), $\rho$ is the target density (for argon, corrected for the nominal density of gas in the target cell), $BH$ is the local density change due to the beam heating the gas cell times the gas expansion due to boiling effects (this correction is not included in the case of \isotope[48][]{Ti}), $V_B$ is the effect of the acceptance and kinematical cuts, and $C_\text{rad}$ is the effect of the radiative corrections and bin center migration.

We used the SIMC spectrometer package~\cite{SIMC} to simulate $(e,e^\prime p)$ events corresponding to our particular kinematic settings, including geometric details of the target cell, radiation correction, and Coulomb effects. SIMC also provided the $V_B$ and $C_\text{rad}$ corrections as in Eq.~\eqref{eq:yield}.  To simulate the distribution of missing energies and momenta of nucleons bound in the argon and titanium nuclei, SIMC was run with a test SF described in detail in the following subsection.

In Table~\ref{tab:ar_energy_levels} we summarize the energies of the shell model states comprising the ground states of \isotope[40][]{Ar} and \isotope[48][]{Ti}. In our analysis, in case two orbitals overlap in $E_m$, we set the energy range for the orbital to be the same, and we assumed the probability of emission of an electron to be the same. Table~\ref{tab:ar_energy_levels} also lists energies derived from previous data sets, as well as the energy used in the calculation of FSI effects according to the model  described in Sec.~\ref{sec:FSI}.

\begin{table}[htp]
    \centering
    \caption{\label{tab:ar_energy_levels}Parametrization of the missing energy distributions of \isotope[40][18]{Ar} and \isotope[48][22]{Ti} assumed in this analysis. The central peak position $E_{\alpha}$, its width $\sigma_{\alpha}$, and the lower (upper) bound on the considered energy range, $E^\alpha_\text{low}$ ($E^\alpha_\text{high}$) are shown for each level $\alpha$. All values are given in units of MeV.}
    \begin{ruledtabular}
    \begin{tabular}{ c d d d d }
     $\alpha$  & \multicolumn{1}{c}{$E_\alpha$} & \multicolumn{1}{c}{$\sigma_\alpha$} &  \multicolumn{1}{c}{$E^\alpha_\text{low}$} & \multicolumn{1}{c}{$E^\alpha_\text{high}$}\\
     \colrule
     & \multicolumn{4}{c}{argon}\\
   \cline{2-5}
       $1d_{3/2}$ & 12.53 & 2 & 8 & 14\\
       $2s_{1/2}$ & 12.93 & 2 & 8 & 14 \\
       $1d_{5/2}$ & 18.23 & 4 & 14 & 20  \\
       $1p_{1/2}$ & 28.0 & 8 & 20 & 45  \\
       $1p_{3/2}$ & 33.0 & 8 & 20 & 45  \\
       $1s_{1/2}$ & 52.0 & 8 & 45 & 70  \\
     \colrule
     & \multicolumn{4}{c}{titanium}\\
     \cline{2-5}
       $1f_{7/2}$ & 11.45 & 2 &  8 & 14  \\
       $2s_{1/2}$ & 12.21 & 2 & 14 & 30 \\
       $1d_{3/2}$ & 12.84 & 2 & 14 & 30 \\
       $1d_{5/2}$ & 15.46 & 4 & 14 & 30\\
       $1p_{1/2}$ & 35.0 & 8 & 30 & 54 \\
       $1p_{3/2}$ & 40.0 & 8 & 30 & 54 \\
       $1s_{1/2}$ & 62.0 & 8 & 53 & 80 \\
    \end{tabular}
    \end{ruledtabular}
\end{table}

SIMC generates events for a broad phase-space, and propagates the events through a detailed model of the electron and proton spectrometers to account for acceptances and resolution effects. Each event is weighted by the $\sigma_{cc1}$ cross section of de~Forest~\cite{deforest} and the SF. The final weighted events do not contain any background. As pointed out above, SIMC does not include FSI corrections other than for the nuclear transparency.

The data yield corrected for the above-mentioned factors is then integrated over $E_{m}$ to get the cross section as function of $p_{m}$. We collected 29.6 (12.5) hours of data on Ar (Ti), corresponding to $\approx$44k (13k) events.

We estimated the background due to accidentals to be 2\% (3\%) for Ar (Ti), performing analysis for each bin of $E_m$ and $p_m$. First, we selected events in T1 trigger in anti-coincidence between the electron and proton arms. This region corresponds to 100 times the nominal coincidence time window width ($\approx$2~ns). Then, we re-scaled the total number of events found to the width of the coincidence peak to obtain a correct estimate of the background events. The background-event distributions were then generated and subtracted bin by bin from the $E_m$ and $p_m$ distributions.


\subsection{Test spectral functions}\label{sec:SF}

The spectral function employed to simulate events in SIMC is based on the simplest implementation of the nuclear shell model,
\begin{equation}\label{eq:testSF}
P(\ve p_m,E_m) = \sum_{\alpha} |\phi_\alpha(\ve p_m)|^2 f_\alpha(E_m-E_\alpha) \ ,
\end{equation}
where the sum runs over all occupied states. In the above equation, $\phi_{\alpha}(\ve p_m)$ is the momentum-space wave function of the state $\alpha$, normalized to unity, and $f_\alpha(E_m-E_\alpha)$ represents the distribution of missing energy peaked at the value $E_\alpha$, reflecting the width of the corresponding state. As a consequence of deviations from this mean-field picture originating from nucleon-nucleon correlations, we expect the Monte Carlo simulations typically to overestimate the data, due to the partial depletion of the shell-model states and to the correlated contribution to the nuclear spectral function.

\begin{figure}[tp]
    \centering
    \includegraphics[width=1.0\columnwidth]{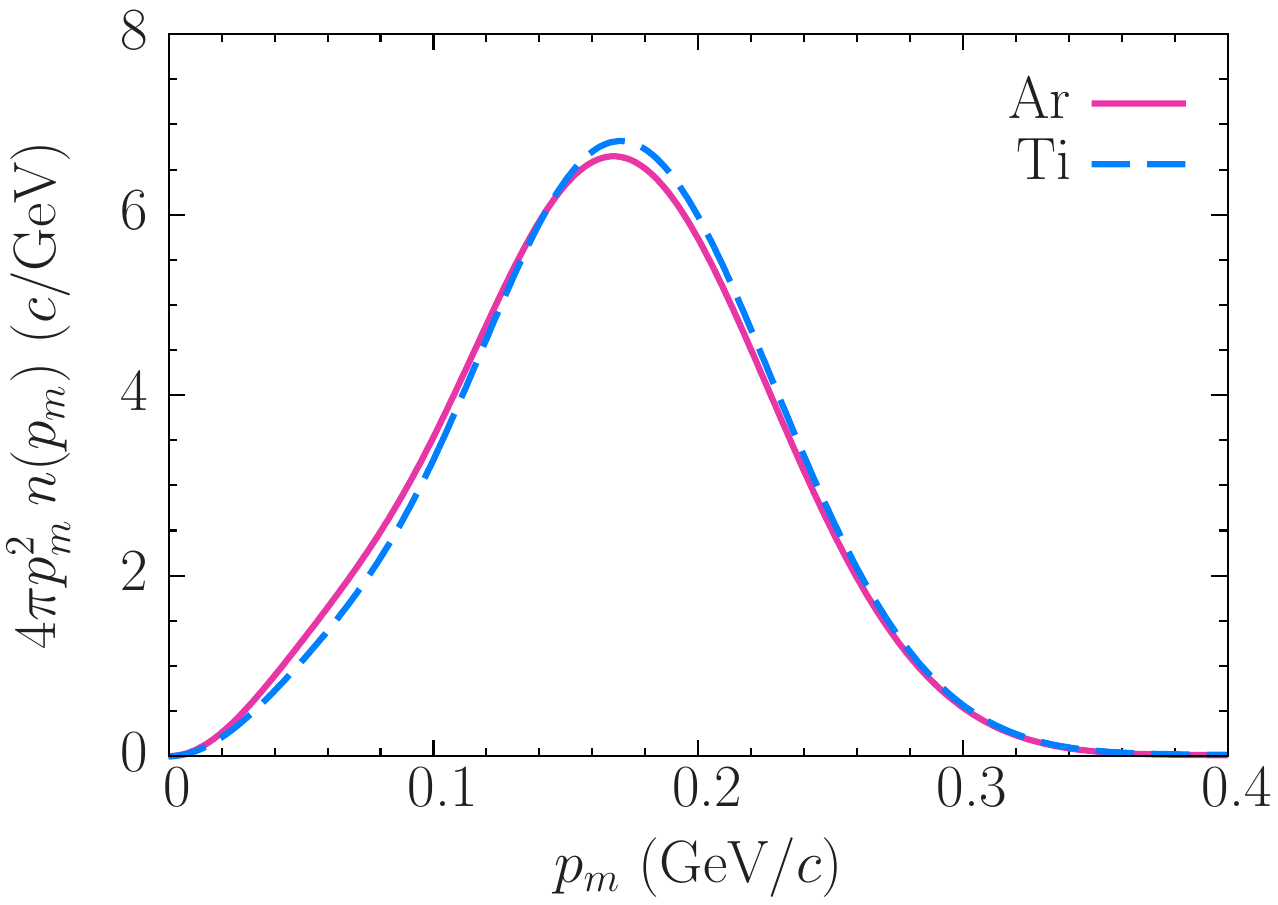}
    \caption{Missing momentum distributions of protons in argon and titanium assumed in this analysis.}
    \label{fig:p_distribution}
\end{figure}

We compared the momentum distribution, defined as
\begin{align}
n(p_m) = \int{P(p_m,E_m) dE_m} ,
\end{align}
obtained using the wave functions of Refs.~\cite{BCSpaper,BCSpaper1}
and Ref.~\cite{DHBpaper}, and found that the differences between them are negligible for both argon and titanium. As shown in Fig.~\ref{fig:p_distribution}, the momentum distributions for argon and titanium also turn out not to differ significantly. This finding suggests that nuclear effects in argon and titanium are similar.

The missing energy distributions are assumed to be Gaussian

\begin{align}
f_\alpha(E_m-E_\alpha)=\frac{1}{\sqrt{2\pi}\sigma_\alpha}\exp\left[-\frac{(E_m-E_\alpha)^2}{2\sigma_\alpha^2}\right].
\end{align}

We obtain the missing energies of the least-bound valence orbital for protons---corresponding to the residual nucleus being left in the ground state, with an additional electron and the knocked-out proton at rest---from the mass difference of the residual system and the target nucleus~\cite{Wang:2017}. These values of missing energy, corresponding to the $1d_{3/2}$ ($1f_{7/2}$) state for \isotope[40][18]{Ar} (\isotope[48][22]{Ti}) in Table~\ref{tab:ar_energy_levels}, are given by
\[
E_{\rm thr}=M_{A-1}+M+m-M_A,
\]
where $m$ stands for the electron mass.

In principle, the energies of other valence levels of \isotope[40][18]{Ar} and \isotope[48][22]{Ti} could be obtained from the excitation spectra of \isotope[39][17]{Cl}~\cite{Chen:2018trb} and \isotope[47][21]{Sc}~\cite{Burrows:2007jwj}. However, the fragmentation of shell-model states induced by long-range correlations makes this information difficult to interpret within the independent-particle model, assumed in Eq.~\eqref{eq:testSF}, because a few spectroscopic lines typically correspond to a given spin-parity state. To overcome this issue and identify the dominant lines, we rely on the spectroscopic strengths determined in past direct pick-up experiments such as $A(\isotope[2][1]{H}, \isotope[3][2]{He})$ for argon~\cite{Mairle:1993asu} and titanium~\cite{Doll:1979}.

The heavily fragmented $1d_{5/2}$ shell~\cite{Mairle:1993asu,Doll:1979}---with over 10, densely packed, spectroscopic lines contributing---can be expected to lend itself well to the approximation by a single distribution of finite width. To determine its peak position, in addition to the experimental data~\cite{Mairle:1993asu,Doll:1979}, we use the theoretical analyses of Refs.~\cite{Warburton:1989zz,Doll:1977cpp} as guidance.

\begin{figure}
    \centering
    \includegraphics[width=1.0\columnwidth]{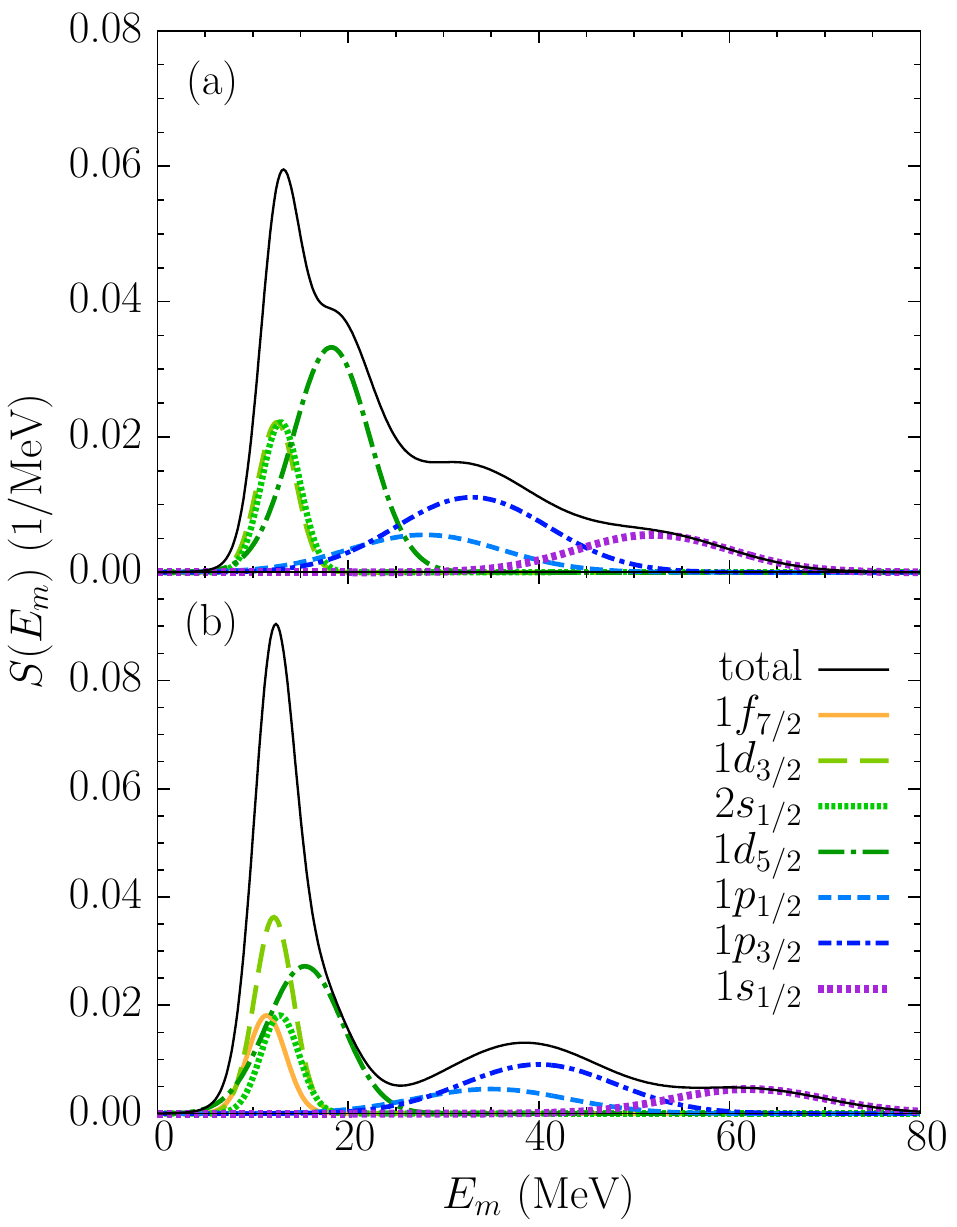}
    \caption{Missing energy distribution of protons in (a) argon and (b) titanium assumed in this analysis.}
    \label{fig:xsec_shell}
\end{figure}

More deeper-lying shells---$1p_{1/2}$, $1p_{3/2}$, and $1s_{1/2}$---were not probed by the past experiments~\cite{Mairle:1993asu,Doll:1979}.
Their $E_\alpha$ values, as well as the widths $\sigma_\alpha$ for all shells, are determined to provide a reasonable description of the missing-energy distributions obtained in this experiment. The resulting parametrization is detailed in Table~\ref{tab:ar_energy_levels}, and presented in Fig.~\ref{fig:xsec_shell}.

%
%
\begin{table}[tb!]
\caption{\label{tab:syst}Contributions to systematical uncertainties for argon and titanium average over all the $E_m$ and $p_m$ bins.}
\begin{ruledtabular}
\begin{tabular}{@{}l c c c@{}}
					&						            	& \multicolumn{1}{c}{Ar}	& \multicolumn{1}{c}{Ti}     \\
\hline
{1.~Total statistical uncertainty} &			        				& 0.53\%   			& 0.78\% \\
{2.~Total systematic uncertainty}  &			        				& 2.75\%   			& 2.39\% \\
\phantom{1. }a.~Beam $x\&y$ offset &						& 0.56\% 				& 0.48\% \\
\phantom{1. }b.~Beam energy & 					    		& 0.10\% 				& 0.10\% \\
\phantom{1. }c.~Beam charge &					    		& 0.30\% 				& 0.30\% \\
\phantom{1. }d.~HRS $x\&y$ offset &				    		& 0.72\% 				& 0.69\% \\
\phantom{1. }g.~Optics (q1, q2, q3) &						& 1.10\% 				& 0.34\% \\
\phantom{1. }h.~Acceptance cut $(\theta,\phi,z)$ & 				& 1.23\% 				& 1.39\% \\
\phantom{1. }i.~Target thickness/density/length & 				& 0.2\% 				& 0.2\% \\
\phantom{1. }j.~Calorimeter \& \v{C}erenkov cut & 				& 0.02\% 				& 0.02\% \\
\phantom{1. }k.~Radiative and Coulomb corr. & 				& 1.00\%				& 1.00\% \\
\phantom{1. }l.~$\beta$ cut		 &				    		& 0.63\%				& 0.48\% \\
\phantom{1. }m.~Boiling effect		&						& 0.70\%				& \multicolumn{1}{c}{\phantom{111}---} \\
\phantom{1. }n.~Cross section model &						& 1.00\%				& 1.00\% \\
\phantom{1. }o.~Trigger and coincidence time cut &				& 0.99\%				& 0.78\% \\

\end{tabular}
\end{ruledtabular}
\end{table}

\section{Uncertainty Analysis}\label{sec:Uncertainties}
The total systematic uncertainty in this analysis was estimated by summing in quadrature the contributions listed in Table~\ref{tab:syst}. We determined the kinematic and acceptance cuts ensuring that there are no dependencies on kinematic variables and input theory model, in this way all uncertainties are uncorrelated bin to bin. All the kinematic and acceptance cuts were varied by the resolution of the variable under consideration. Except for the transparency corrections, the MC used to evaluate those uncertainties did not contain effects due to FSI, such as a quenching of the strength of the cross section and a modification of the kinematic of the outgoing particles. {\it A priori} the MC simulation could depend on the underlying theoretical model. However, we repeated the analysis of systematic uncertainties varying its ingredients, and did not observe any substantial variations of the obtained results. 
%
As the obtained results depend on the Monte Carlo calculation, it is important to estimate uncertainties resulting from its inputs. To determine the uncertainties related to the target position, we performed the simulation with the inputs for the beam's and spectrometer's $x$ and $y$ offsets varied within uncertainties, and we recomputed the optical transport matrix varying the three quadrupole magnetic fields, one at the time. Each of these runs was compared to the reference run, and the corresponding differences were summed in quadrature to give the total systematic uncertainty due to the Monte Carlo simulation.
\begin{figure}[bt]
\subfloat[Ar]{\includegraphics[width=1.0\columnwidth]{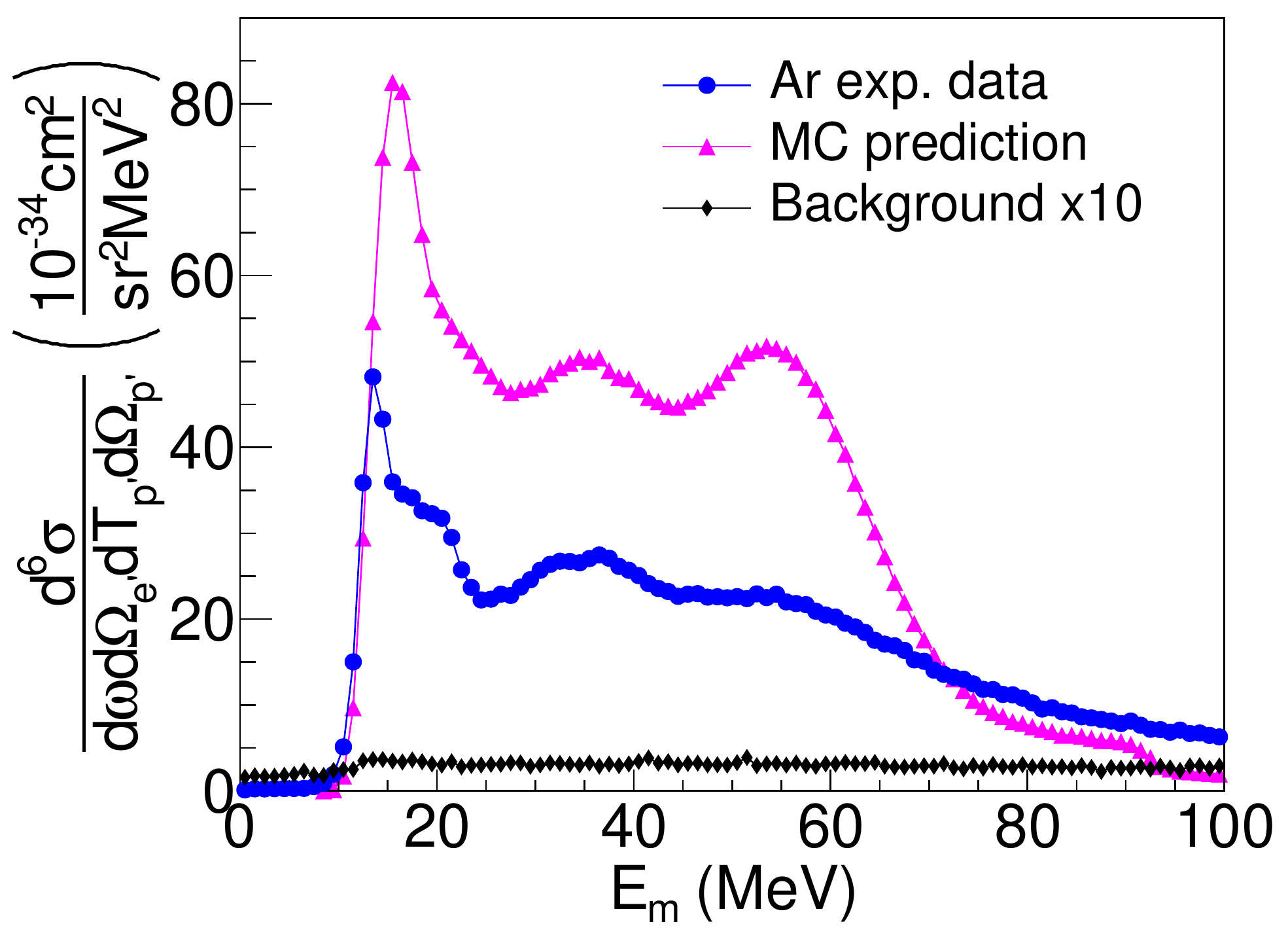}}\qquad
\subfloat[Ti]{\includegraphics[width=1.0\columnwidth]{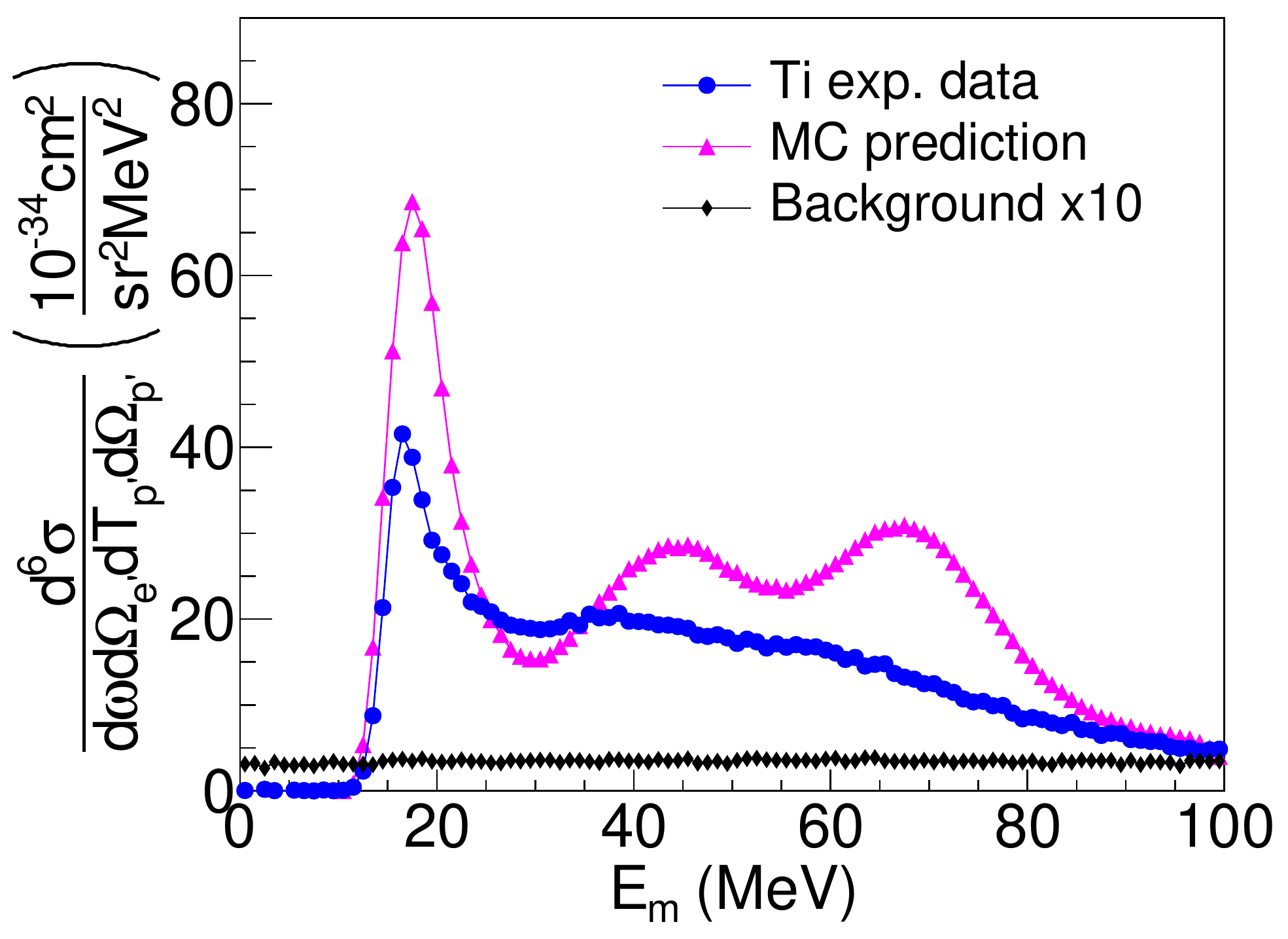}}\qquad
\caption{\label{fig:sigma_tot_em}Six-fold differential cross section as a function of missing energy for argon ((a) panel) and titanium ((b) panel). The background estimate (line connecting the experimental data points)  is multiplied by 10 for purpose of presentation. The MC predictions, based on the mean-field SF, include a correction for the nuclear transparency, while  other FSI effects are not accounted for. }
\end{figure}
\begin{figure}[bt]
\centering
\subfloat[Ar\label{sigma_tot_pm_ar}]{\includegraphics[width=1.0\columnwidth]{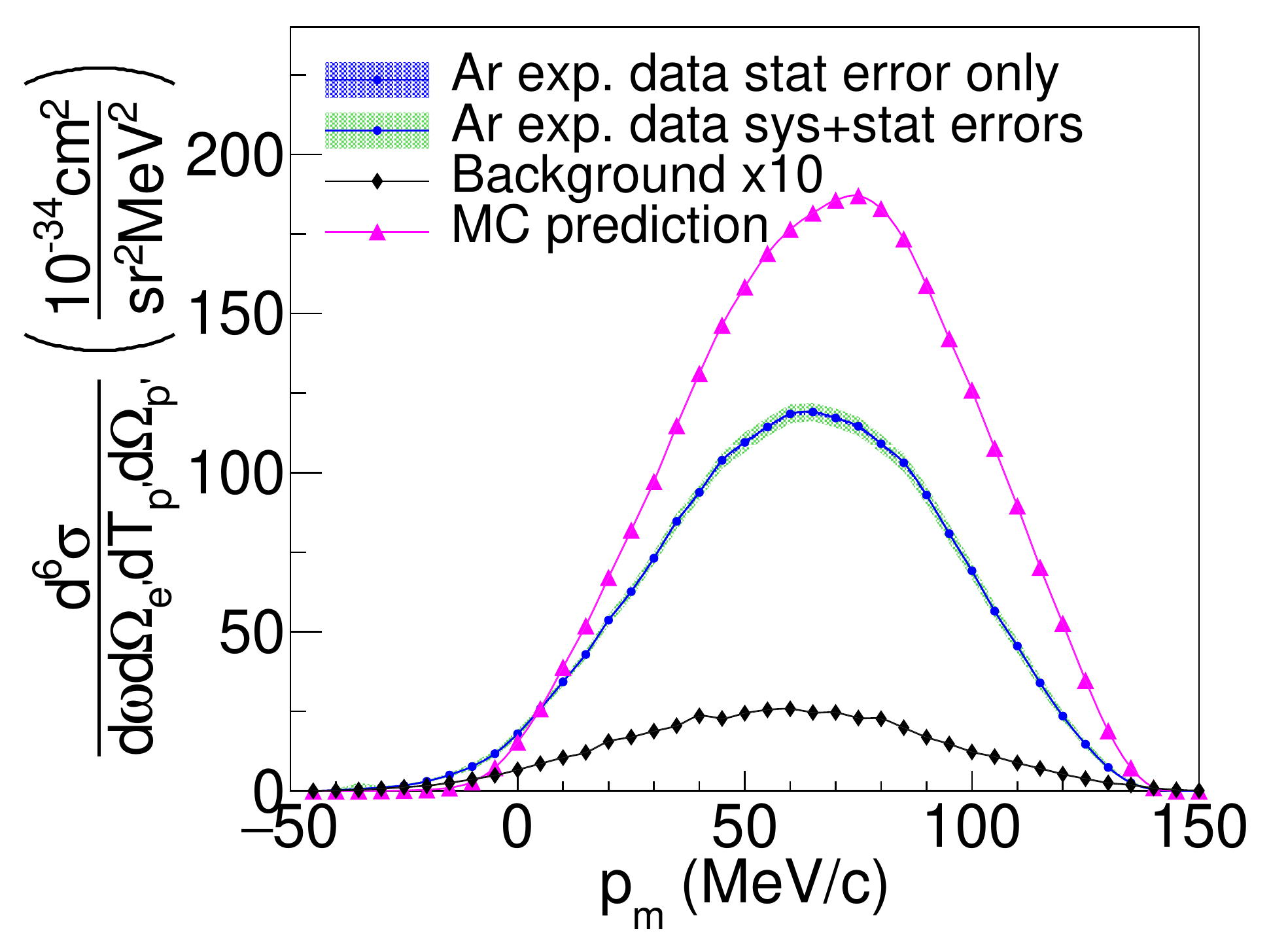}}\qquad
\subfloat[Ti\label{sigma_tot_pm_ti}]{\includegraphics[width=1.0\columnwidth]{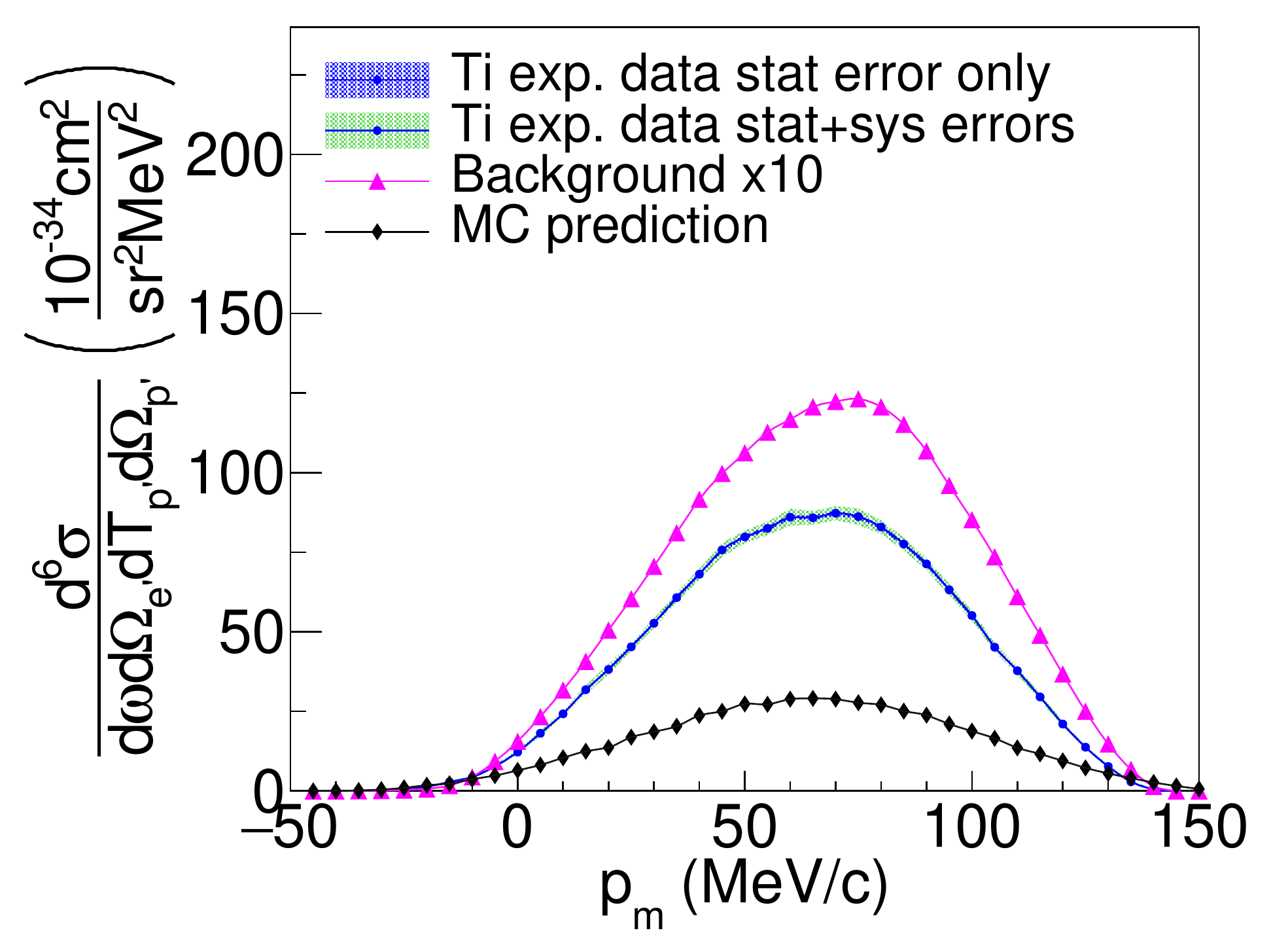}}\qquad
\caption{\label{fig:sigma_tot_pm}Same as Fig.~\ref{fig:sigma_tot_em} but for the cross section as a~function of missing momentum. The inner (outer) uncertainty bands correspond to statistical (total) uncertainties.}
\end{figure}
The uncertainties related to the calorimeter and \v{C}e{\-}ren{\-}kov detectors were determined by changing the corresponding cut by a small amount and calculating the difference with respect to the nominal yield value. The uncertainty due to the acceptance cuts on the angles was calculated using the same method. We included an overall fixed uncertainty for both the beam charge and beam energy, as in the previous work on C, Ti, Ar, and Al~\cite{Dai:2018gch,Dai:2018xhi,Murphy:2019wed}. We evaluated the systematic uncertainties related to the trigger efficiency by determining variations across multiple runs, as well as by applying different acceptance cuts. A fixed uncertainty was assigned to take care of those variations.

The time-coincidence cut efficiency, as other acceptance cuts, was evaluated by changing the cut by $\pm \sigma$.

SIMC generates events including the effects from radiative processes: vacuum polarization, vertex corrections, and internal bremsstrahlung. External radiative processes refer to electrons losing energy while passing through material in the target. Radiative correction in SIMC are implemented following the recipe of Dasu~\cite{Dasu}, using the Whitlow's approach~\cite{Tsai:1969,Whitlow:1990}. We considered a fixed 1\% uncertainty due to the theoretical model for the radiative corrections over the full kinematic range as in our previous work. We generated different MC where the radiative corrections were re-scaled by $\sqrt{(Q^2)}/2$, $Q^2$ being the four-momentum transfer squared, and re-analyzed the data and looked for variations. Coulomb corrections were included in the local effective momentum approximation~\cite{Aste:2005wc}. A 10\% uncertainty associated with the Coulomb potential was included as systematic uncertainty. Finally, we included a target thickness uncertainty and an uncertainty due to the boiling effect correction~\cite{Santiesteban:2018qwi}.

The measured and MC predicted differential cross sections ${d^6 \sigma }/{ d\omega d\Omega_e dp d\Omega_p}$ are presented in Fig.~\ref{fig:sigma_tot_em} as a function of $E_m$ and in Fig.~\ref{fig:sigma_tot_pm} as a function of $p_m$, integrated over the full range of $E_m$, for \isotope[40][]{Ar} (panel (a)) and \isotope[48][]{Ti} (panel (b)) targets.

The MC simulation clearly overestimates the extracted cross sections. As the nuclear model underlying the simulation neglects the effects of FSI other than the nuclear transparency and all correlations between nucleons, this difference is by no means surprising. Both FSI and partial depletion of the shell-model states require further studies, base on all five datasets collected by the JLab E12-14-012 experiment, which will be reported elsewhere.


\subsection{Final state interactions}\label{sec:FSI}

Within DWIA, FSI between the outgoing proton and the spectator nucleons are described by a complex, energy dependent, phenomenological optical potential (OP).
The OPs available for calculations were determined by fitting a set of elastic proton-nucleus scattering data for a range of target nuclei and beam energies. Different parametrizations, yielding equivalently good descriptions of the data, can give differences and theoretical uncertainties when ``equivalent'' OPs are used in kinematical regions for which experimental data are not available, or when they are extended to inelastic scattering and to calculation of the cross section of different nuclear reactions.

Nonrelativistic and relativistic OPs are available for $(e,e^\prime p)$ calculations within nonrelativistic and relativistic DWIA frameworks. However, nonrelativistic phenomenological OPs are available for energies not larger than 200~MeV. It is generally believed that above $\approx$180~MeV the Schr\"{o}dinger picture of the phenomenological OP should be replaced by a Dirac approach, and a relativistic OP should be used. In Ref.~\cite{meu01a}, it was shown that in $(e,e^\prime p)$ reactions the differences between the nonrelativistic and relativistic DWIA results depend on kinematics and increase with the outgoing proton energy, and for proton energies above 200~MeV a relativistic calculation is necessary.

\begin{figure}[bt]
    \centering
    \includegraphics[width=1.0\columnwidth]{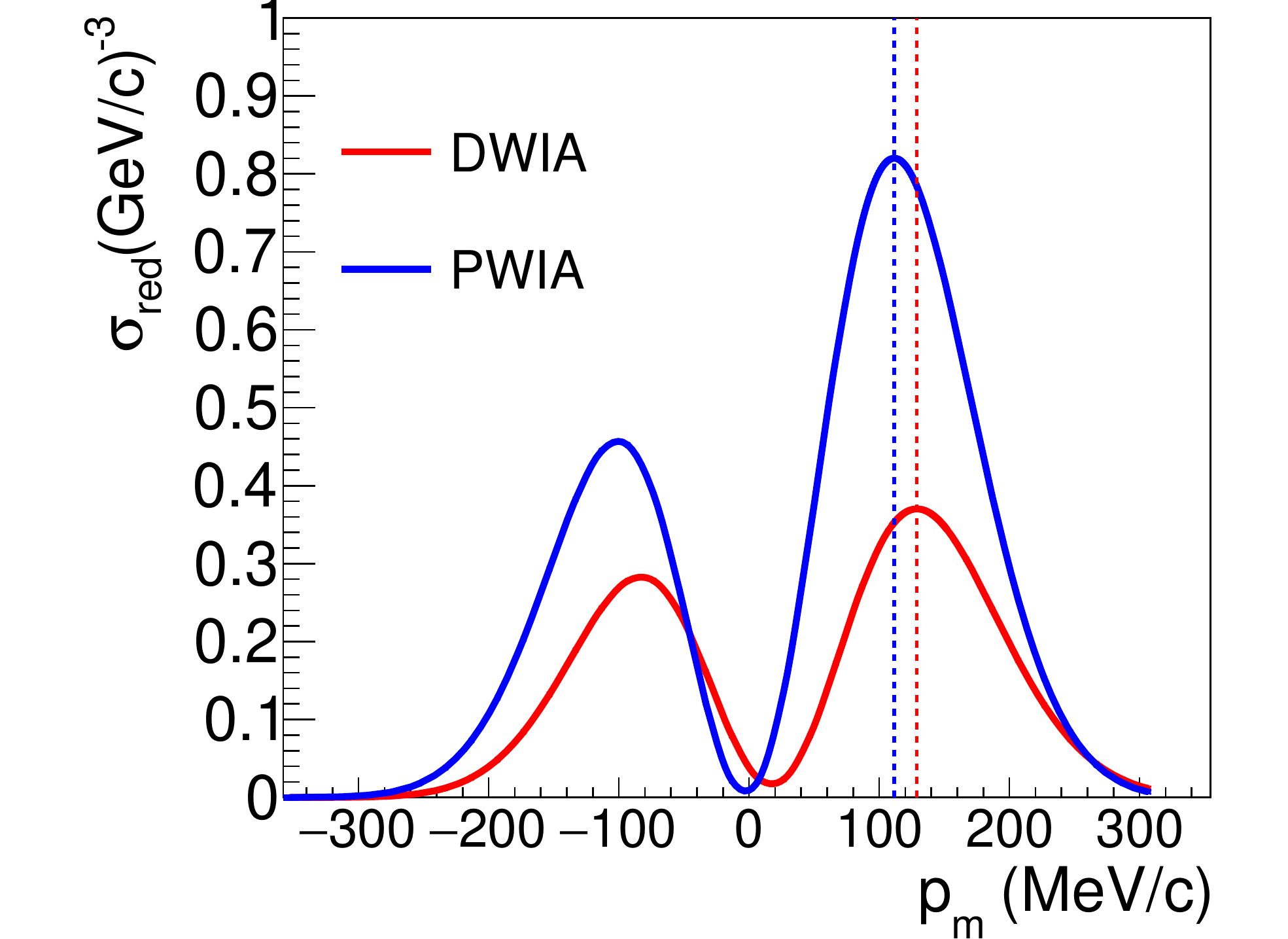}
    \caption{\label{fig:redxs_1p12}Reduced cross section as a function of missing momentum for the $1p_{1/2}$ proton knockout from argon. We compare the PWIA and DWIA results obtained for the parallel kinematics considered in this analysis.}
\end{figure}

We have used the so-called ``democratic''(DEM) relativistic OP~\cite{coo09}, obtained from a global fit to over 200 sets of elastic proton-nucleus scattering data, comprised of a broad range of targets, from helium to lead, at energies up to 1,040~MeV.

An example of the comparison between PWIA and DWIA results is given in Fig.~\ref{fig:redxs_1p12}, where the reduced cross section as a function of $p_m$ is displayed for proton knockout from the $1p_{1/2}$ argon orbital. Calculations are performed within the relativistic model of Ref.~\cite{meu01a} for the parallel kinematics of the present experiment.
Positive and negative values of $p_m$ indicate, conventionally, cases in which $|\q|<|\p^\prime|$ and $|\q|>|\p^\prime|$, respectively. The reduction and the shift produced in the reduced cross section by FSI in the DWIA calculation can be clearly seen.

The two dashed lines drawn in the region of positive $p_m$ of the figure indicate the value of $p_m$ corresponding to the peaks of the DWIA and PWIA  reduced cross sections. We use the distance between the two dashed lines as a measure of the shift produced by FSI.

The reduction of the calculated cross section produced by FSI can be measured by the DWIA/PWIA ratio, which is defined here as the ratio of the integral over $p_m$ of the DWIA and PWIA reduced cross sections. Both the shift and the DWIA/PWIA ratios are computed separately for the positive and negative $p_m$ regions.

\begin{table}
    \centering
    \caption{\label{tab:opt_pos}Shifts between the reduced DWIA and PWIA cross sections, and the DWIA to PWIA cross-section ratios, obtained for proton knockout from various argon orbital using different optical potentials: DEM~\cite{coo09}, EDAD3~\cite{coo93}, and EDAD1~\cite{coo93}. All results are calculated for $p_m > 0$.}
    \begin{ruledtabular}
    \begin{tabular}{c d d d d d d}
    \multirow{2}{*}{Orbital}& \multicolumn{3}{c}{Shift (MeV/$c$)}& \multicolumn{3}{c}{{DWIA}/{PWIA}}\\
         \cline{2-4}\cline{5-7}\\[-8pt]
        &\multicolumn{1}{c}{EDAD1} &  \multicolumn{1}{c}{EDAD3}& \multicolumn{1}{c}{DEM}& \multicolumn{1}{c}{EDAD1} &  \multicolumn{1}{c}{EDAD3} & \multicolumn{1}{c}{DEM}\\
        \colrule
          $1d_{3/2}$ & 1.5  & -2.0 &1.5  & 0.58 & 0.57 &0.58 \\
          $2s_{1/2}$ & 8.0  & 7.0  &8.0  &0.78 & 0.78  &0.78\\
          $1d_{5/2}$ &-2.0 & -6.5& -3.0 &0.57 & 0.57 &0.58\\
          $1p_{1/2}$ & 12.5 & 9.0 &12.5  &0.43 & 0.39 &0.42\\
          $1p_{3/2}$ & 9.5  & 5.0  &9.0  &0.47 & 0.44 &0.46\\
          $1s_{1/2}$ & 13.0 & 10.0 &13.0 &0.42 & 0.38 &0.41\\
    \end{tabular}
    \end{ruledtabular}
\end{table}

The theoretical uncertainty of the shift and the reduction produced by FSI has been evaluated investigating the sensitivity of the DWIA and PWIA results to different choices of the theoretical ingredients of the calculation.

The uncertainty due to the choice of the OP has been evaluated by comparing the results obtained with the DEM and other energy-dependent and atomic-number dependent relativistic OPs, referred to as EDAD1 and EDAD3~\cite{coo93} .
The shift and the DWIA/PWIA ratio in the positive $p_m$ region, computed for proton knock out from various argon orbitals using the DEM, EDAD1, and EDAD3 potentials  are reported in Table~\ref{tab:opt_pos}. The results indicate a slight dependence of FSI effects on the choice of OP.

\begin{figure}
    \centering
    \includegraphics[width=1.0\columnwidth]{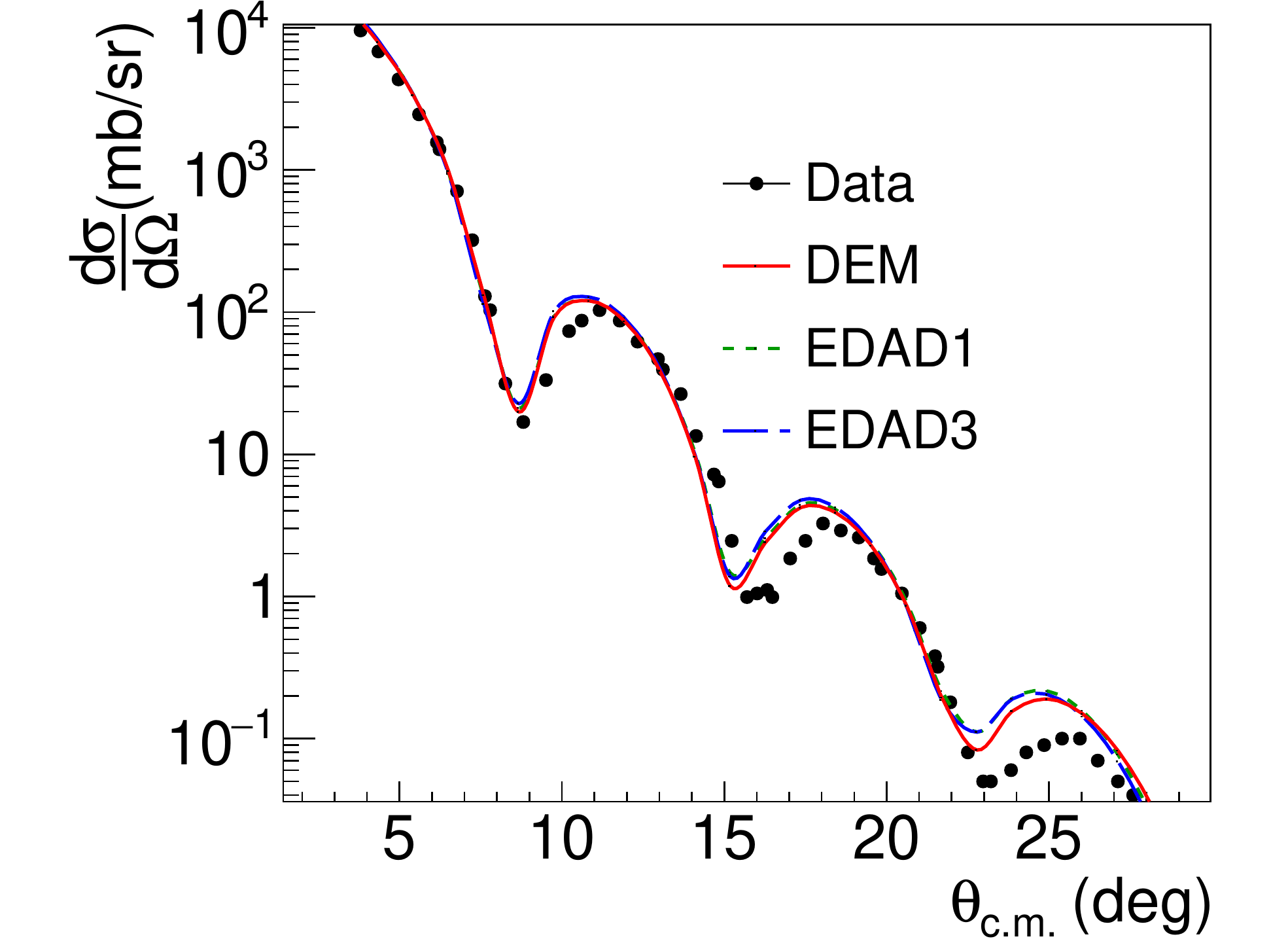}
    \caption{\label{fig:xsec_theta}Differential cross section for elastic proton scattering on \isotope[40][]{Ar} at 0.8~GeV as a function of scattering angle. Results for the DEM, EDAD1, and EDAD3 optical potentials, which turn out to almost completely overlap, are compared with the experimental data~\cite{elAlamos}.}
\end{figure}

Note that the three OPs were determined by a fitting procedure of elastic proton scattering data over a wide range of nuclei, which, however, did not include  argon. This means that the ability of the phenomenological OPs to describe elastic proton scattering data  on argon is not guaranteed. A test of this ability is presented in Fig.~\ref{fig:xsec_theta}, where the \isotope[40][]{Ar}$(p,p^\prime)$ cross section calculated at 0.8~GeV with the three OPs is compared  to the corresponding experimental cross section obtained using the HRS of the Los Alamos Meson Physics Facility~\cite{elAlamos}. The results of the three OPs largely overlap, and their agreement with the experimental cross section, although not perfect, is more than reasonable, in particular if we consider that it has not been obtained from a fit to the data.

In the relativistic DWIA and PWIA calculations different current conserving (cc) expressions of the one-body nuclear current operator can be adopted. 
The different expressions are equivalent for on-shell nucleons, while differences can arise for off-shell nucleons. For all the results that we have presented until now, and as a basis for the present calculations, we have adopted the cc1 prescription~\cite{deforest}. We note that, historically, the cc1 cross section has been often used to obtain the reduced cross section from the experimental and theoretical cross section. The impact of using a different cross section\textemdash such as the cc2 model of Ref.~\cite{deforest}\textemdash in the determination of the spectral function will be discussed in future analysis.

We have also checked that the differences obtained using different proton form factors in the calculation of the nuclear current are always negligible in the kinematic situation of the present experiment.

The bound proton states adopted in the calculations are self-consistent Dirac-Hartree solutions derived within a relativistic mean field approach using a Lagrangian containing $\sigma$, $\omega$, and $\rho$ mesons, with medium dependent parametrizations of the meson-nucleon vertices that can be more directly related to the underlying microscopic description of nuclear interactions~\cite{BCSpaper,BCSpaper1}. Pairing effects have been included carrying out Bardeen-Cooper-Schrieffer (BCS) calculations. The theoretical uncertainties on the shift and the DWIA/PWIA ratio due to the use of wave functions obtained with a  different description of pairing, based on the relativistic Dirac-Hartree-Bogoliubov (DHB) model~\cite{DHBpaper},  turn out to be negligible.

In our analysis we assumed the missing energy distribution for each of the orbitals in \isotope[40][]{Ar} and \isotope[48][]{Ti} as shown in Fig.~\ref{fig:xsec_shell}. The lower and upper energy bounds assumed in the DWIA analysis of FSI are given for each orbital in Table~\ref{tab:ar_energy_levels}. 
The FSI correction has been applied event by event in both the missing energy and missing momentum distributions. We applied different corrections for events with $|\q|<|\p^\prime|$ and $|\q|>|\p^\prime|$, according to the theoretical predictions mentioned before. For each event, we used the reconstructed energy and momentum of both electron and proton to determine the orbital involved in the primary interaction. Then, we applied the FSI correction, based on the $p_m$ sign. For orbitals that overlap we use a simple prescription to determine the most probable orbital from which the electron was emitted, as described in Sec.~\ref{subsec:Analysis}.

%
%
%
\section{Differential cross section comparison}\label{sec:comparison}
%

\begin{figure}[htp!]
\centering
\subfloat[$\quad0<E_m < 27$~MeV]{\includegraphics[width=1.0\columnwidth]{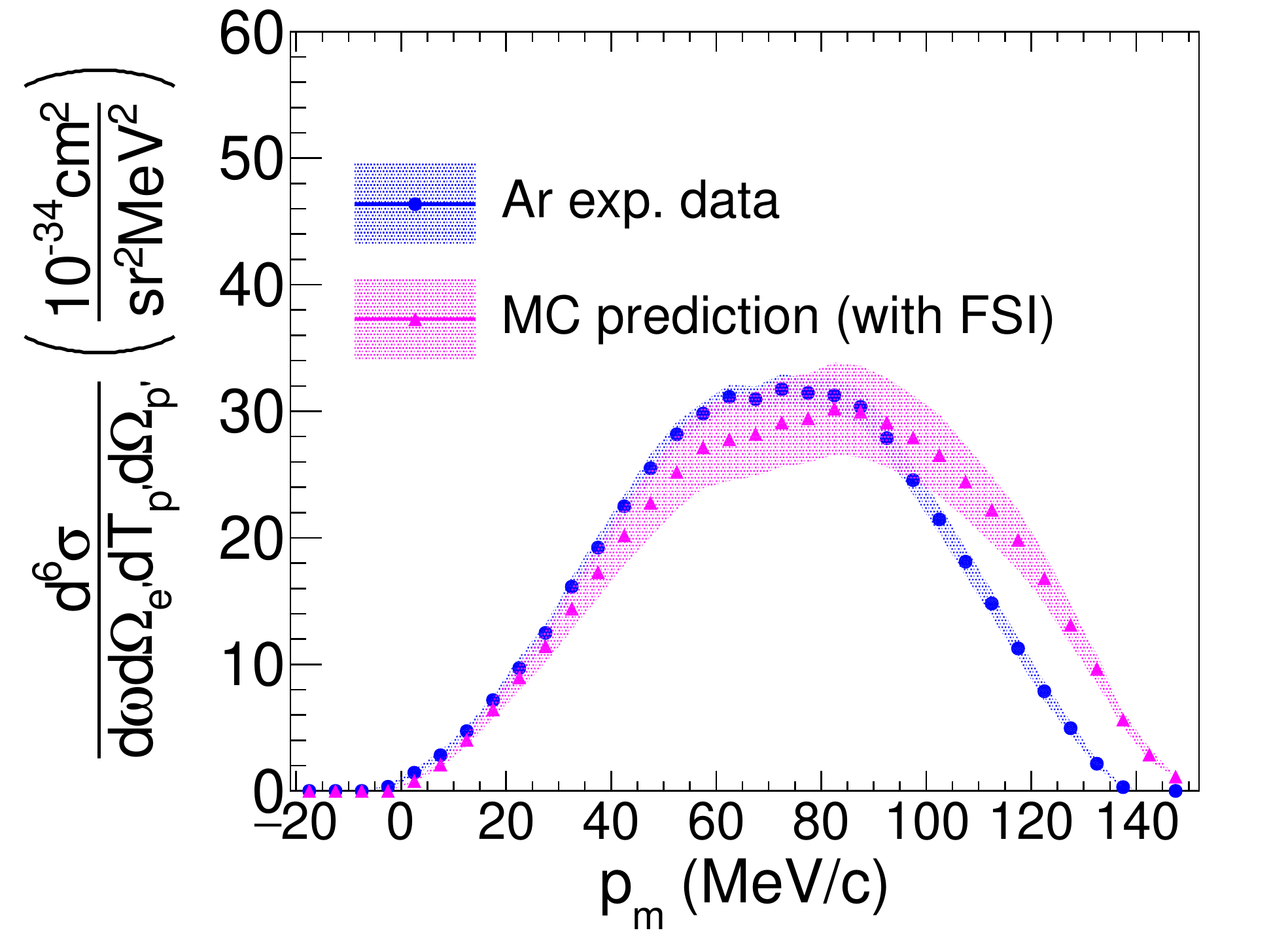}}\qquad
\subfloat[$\quad27< E_m < 44$~MeV]{\includegraphics[width=1.0\columnwidth]{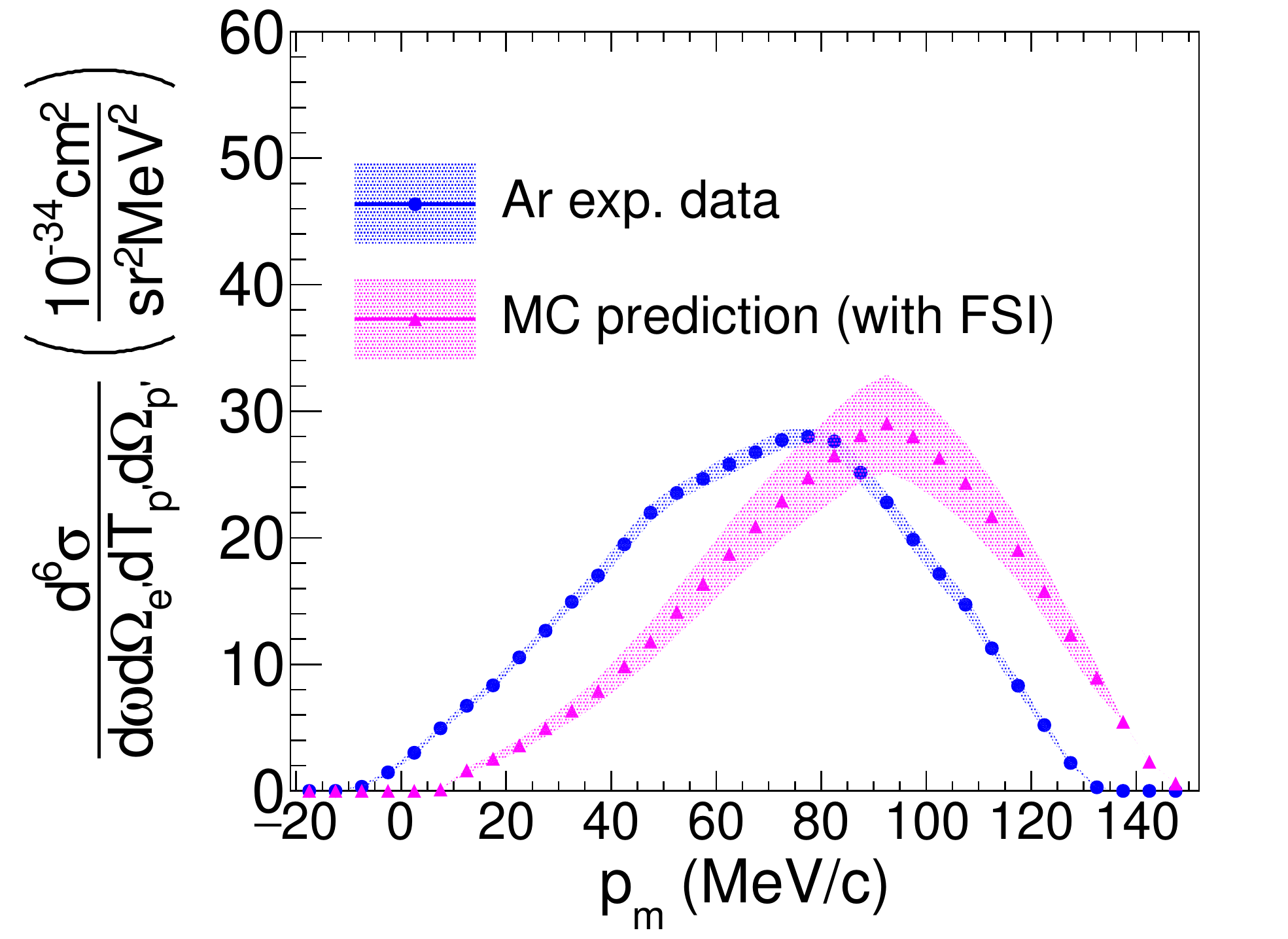}}\qquad
\subfloat[$\quad44 < E_m < 70$~MeV]{\includegraphics[width=1.0\columnwidth]{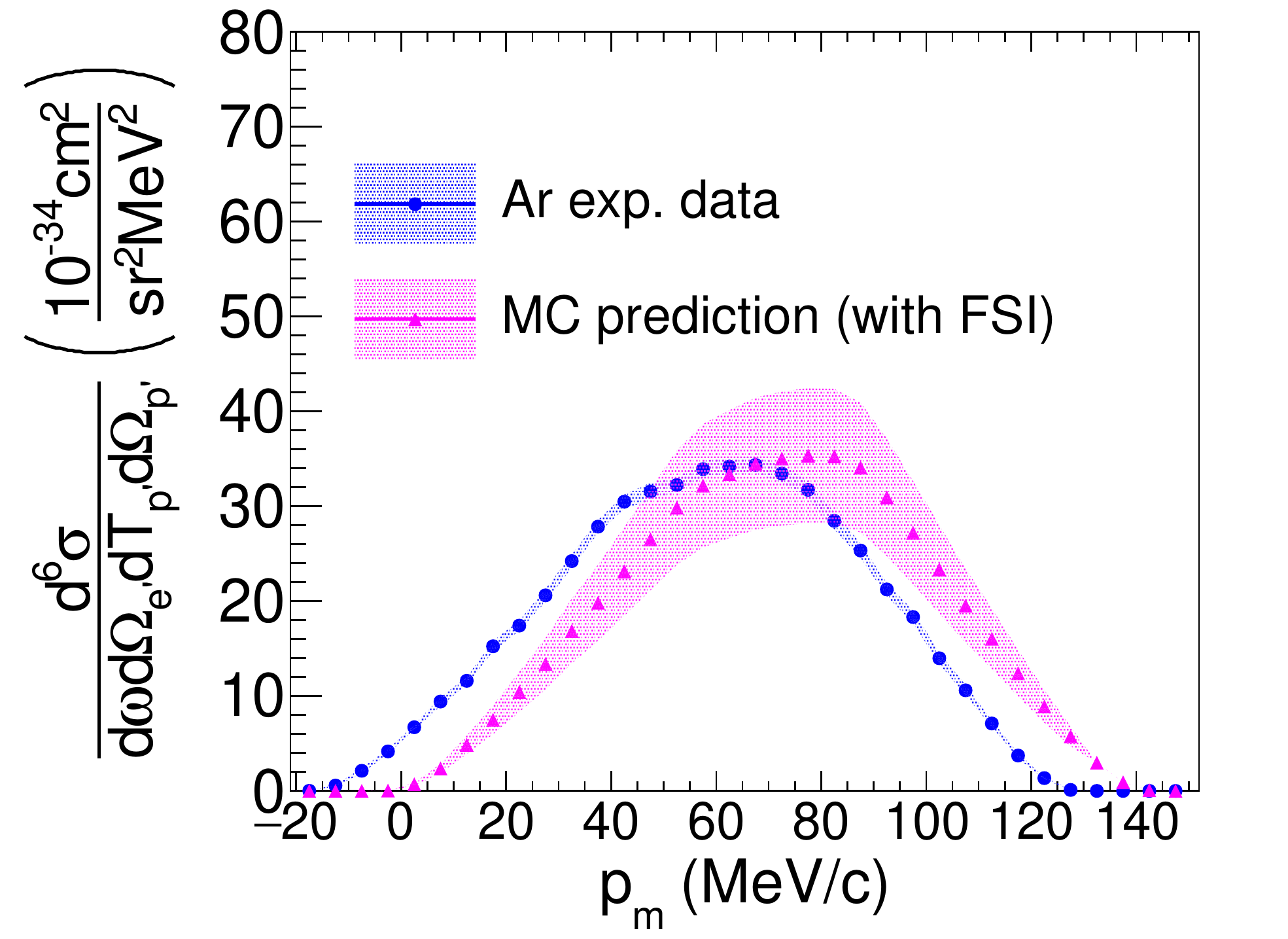}}\qquad
\caption{\label{fig:redxs_pm_Ar}Six-fold differential cross section for argon as a function of missing momentum integrated over different ranges of missing energy. The background estimate is multiplied by 10 for presentation. The MC predictions, based on the mean-field SF, include the full FSI corrections.}
\end{figure}

\begin{figure}[htp!]
\centering
\subfloat[$\quad0 <E_m < 30$~MeV]{\includegraphics[width=1.0\columnwidth]{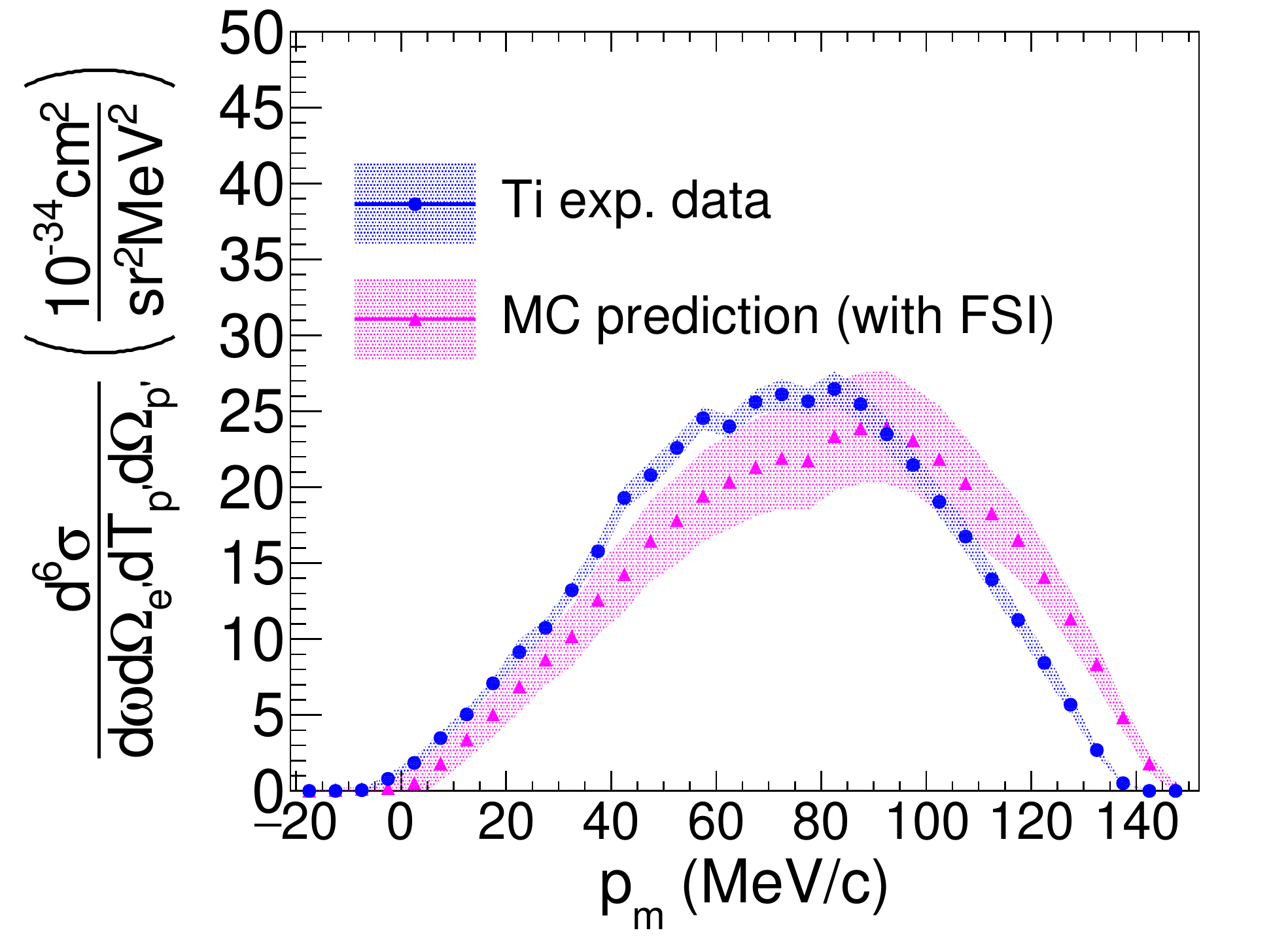}}\qquad
\subfloat[$\quad30< E_m < 54$~MeV]{\includegraphics[width=1.0\columnwidth]{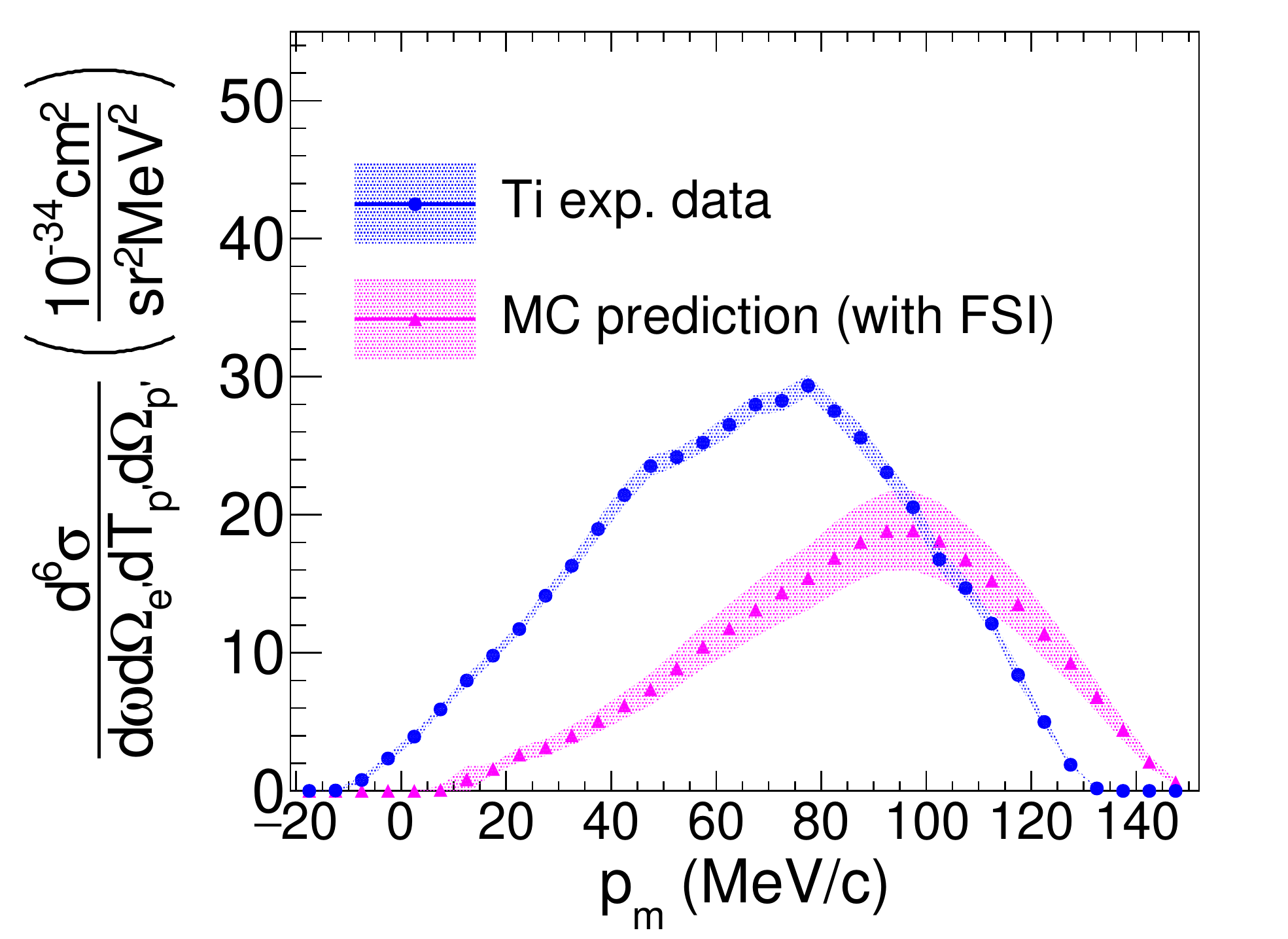}}\qquad
\subfloat[$\quad54 < E_m < 90$~MeV]{\includegraphics[width=1.0\columnwidth]{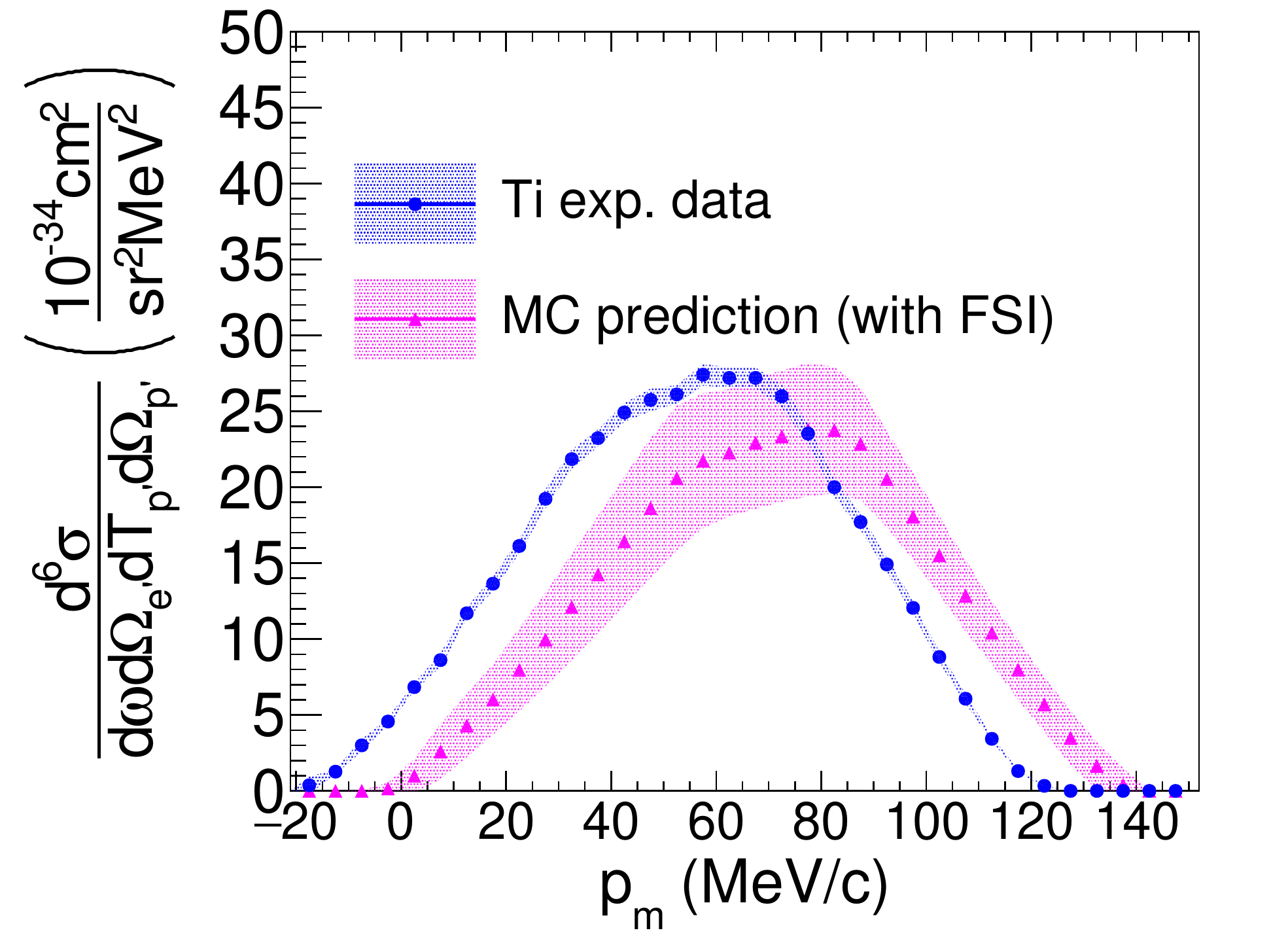}}\qquad
\caption{\label{fig:redxs_pm_Ti}Same as Fig.~\ref{fig:redxs_pm_Ar} but for titanium.}
\end{figure}

Figures~\ref{fig:redxs_pm_Ar} and~\ref{fig:redxs_pm_Ti} show a comparison between the measured differential cross sections of \isotope[40][]{Ar} and \isotope[48][]{Ti} and the MC predictions including full FSI corrections, plotted  as a function of $p_m$ for three different ranges of $E_m$. The missing energy regions  for \isotope[40][]{Ar}  (\isotope[48][]{Ti}) are: $E_m < 27$~MeV ($E_m < 30$~MeV), $27< E_m < 44$~MeV ($30< E_m < 54$~MeV ) and $44 < E_m < 70$~MeV ($54 < E_m < 90$~MeV).

We estimated the background to be of the order 2\% for \isotope[40][]{Ar} and 3\% for \isotope[48][]{Ti}.
The MC systematic uncertainties from FSI are estimated by varying the following ingredients of the model:
\begin{itemize}
    \item[(i)] the optical potential (DEM, EDAD1, or EDAD3);
    \item[(ii)] the pairing mechanism underlying the determination of the wave functions (the default BCS model~\cite{BCSpaper,BCSpaper1} or the DHB model~\cite{DHBpaper});
    \item[(iii)] the parametrization of the  nucleon form factors. 
\end{itemize}
The total systematic uncertainty is obtained by adding in quadrature all the variations, and including an overall uncertainty of the theoretical model of 15\%.


A prominent feature of both Figs.~\ref{fig:redxs_pm_Ar} and~\ref{fig:redxs_pm_Ti} is that the agreement between data and MC predictions including FSI, which turns out to be quite good in the region of low missing energies, becomes significantly worse at larger $E_m$. This behavior can be explained considering that, according to the shell-model picture employed in MC simulations, missing energies $E_m > 27$ MeV correspond to proton knockout from the deeply bound $1p_{1/2}$, $1p_{3/2}$, and $1s_{1/2}$ states. 

As discussed in Sec.~\ref{sec:SF}, the energies and widths of these states are only estimated, and not determined from experimental data. Underestimating the widths and the associated overlaps of energy distributions would imply a smaller value for the differential cross section and a shift in the $p_m$ distribution between data and MC. We have tested this hypothesis by varying the width of the high-energy states in the test SF and redoing our full analysis, and noticed an improved agreement between data and MC. 

More generally, it has to be kept in mind that a clear identification of single particle states in interacting many-body systems\textemdash ultimately based on Landau theory of normal Fermi liquids\textemdash is only possible in the vicinity of the Fermi surface, corresponding to the lowest value of missing energy, see, e.g., Ref.~\cite{BF:book}.  An accurate description of the data at large missing energy will require a more realistic model of the nuclear spectral function, taking into account dynamical effects beyond the mean-field approximation, notably nucleon-nucleon correlations, leading to the appearance of protons in continuum states.
%
%

\section{Summary and Conclusions}\label{sec:Summary}
In this paper, we report the first results of the analysis of $(e,e^\prime p)$ data at beam energy $E_e = 2.222$~GeV an electron scattering angle $\theta_e=21.5$ deg, collected in JLab Hall A by the E12-14-012 experiment using Ar and Ti targets. The measured differential cross sections are presented as a function of missing energy and missing momentum, and compared to the predictions of a MC simulation in which the effects of FSI are described within DWIA.

We were able to select coincidence events between the electron and proton spectrometers with high efficiency and low systematic uncertainties. The level of background and systematic uncertainties turned out to be below 4\%, in line with the goals listed in the original JLab E12-14-012 proposal~\cite{Benhar:2014nca}. Overall, the comparison between the data and results of MC simulations, 
carried  out over the lowest missing energy range  $0 < E_m < 30$~MeV and missing momentum covered by our measurements appears satisfactory. The larger discrepancies observed at the larger missing energies such as $30 < E_m< 44$~MeV  re likely to be ascribable to the limitations of the theoretical model based on the mean-field approximation, employed in MC event generation, which is long known to be inadequate to describe the dynamics of deeply bound nucleons~\cite{Benhar_NPN}. Understanding these discrepancies at quantitative level will require the inclusion of reaction mechanisms beyond DWIA, such as multi-step processes and multi-nucleon emission triggered by nucleon-nucleon correlations.

The missing energy spectra obtained from our analysis contain valuable new information on the internal structure and dynamics of the nuclear targets, encoded in the positions and widths of the observed peaks.  

The determination of these spectra particularly for deep-lying hole excitations is, in fact, a first step towards the derivation of the spectral functions for medium-mass nuclei, such as Ar and Ti, within the framework of LDA, that represents the ultimate aim of our experiment.

The Ar and Ti measurements discussed in this article, providing the first $(e,e^\prime p)$ data in the kinematical range relevant to neutrino experiments---most notably DUNE---comprises the first of five datasets collected by the JLab E12-14-012 experiment. The combined analysis of all data, which is currently under way, will provide information of unparalleled value for the development of realistic nuclear models, and will allow the extraction of Ar and Ti spectral functions. 
%


\begin{acknowledgments}
\par We acknowledge the outstanding support from the Jefferson Lab Hall A technical staff, target group and Accelerator Division. This experiment was made possible by Virginia Tech, the National Science Foundation under CAREER grant No. PHY$-$1352106 and grant No. PHY$-$1757087. This work was also supported by the DOE Office of Science, Office of Nuclear Physics, contract DE-AC05-06OR23177, under which Jefferson Science Associates, LLC operates JLab, DOE contract DE-FG02-96ER40950, DE-AC02-76SF00515, DE-SC0013615 and by the DOE Office of High Energy Physics, contract DE-SC0020262.
\end{acknowledgments}
\newpage
%

\end{document}